\documentclass[acmsmall,screen]{acmart}

\usepackage[utf8]{inputenc}

\usepackage{dsfont}
\usepackage{xcolor}
\usepackage{amsmath}
\usepackage{tcolorbox}
\usepackage{textcomp}
\usepackage{wasysym}
\usepackage{stmaryrd}
\usepackage{caption}
\usepackage{subcaption}
\usepackage{enumitem}
\usepackage{calc}
\usepackage{pifont}
\usepackage{tikz}
\usetikzlibrary{arrows, arrows.meta, calc, positioning, quotes, shapes, graphs,
  shadows}

\AtBeginDocument{%
  \providecommand\BibTeX{{%
    \normalfont B\kern-0.5em{\scshape i\kern-0.25em b}\kern-0.8em\TeX}}}

\setcopyright{rightsretained}
\acmDOI{10.1145/3674636}
\acmYear{2024}
\acmJournal{PACMPL}
\acmVolume{8}
\acmNumber{ICFP}
\acmArticle{247}
\acmMonth{8}
\acmSubmissionID{icfp24main-p45-p}
\received{2024-02-28}
\received[accepted]{2024-06-18}

\definecolor{highlight}{gray}{0.9}
\definecolor{type}{HTML}{AAEEFF}
\definecolor{example}{HTML}{FFAAEE}
\definecolor{sketch}{HTML}{EEFFAA}
\definecolor{type+example}{HTML}{BBAAFF}
\definecolor{example+sketch}{HTML}{FFBBAA}
\definecolor{sketch+type}{HTML}{AAFFBB}
\definecolor{type+example+sketch}{HTML}{EEEEEE}

\tikzset{between/.style args={#1 and #2}{at = ($(#1)!0.5!(#2)$)}}
\tikzset{arr/.style={rounded corners=1.5mm,-latex}}
\tikzset{spec/.style={draw,rounded
  corners=4mm,align=center,minimum height=8mm,inner xsep=4mm}}
\tikzset{stacked/.style=
  {double copy shadow={opacity=1,shadow xshift=0.5mm,shadow yshift=-0.6mm}}}

\newtheorem{definition}{Definition}

\newcommand{\matchlength}[2]{\makebox[0pt][l]{#1}\phantom{#2}}

\newcommand{\xdownarrow}[1]{%
  {\left\downarrow\vbox to #1{}\right.\kern-\nulldelimiterspace}
}
\newcommand{\shape}[1]{\hat{#1}}
\newcommand{\pos}[1]{\mathring{#1}}

\newcommand{\K}{\text{K}}

\newcommand{\Nat}{\mathds{N}}
\newcommand{\Fin}{\textit{Fin}}
\newcommand{\M}{\mathds}
\newcommand\doubleplus{\mathbin{+\mkern-6mu+}}
\makeatletter
\newsavebox{\@brx}
\newcommand{\llangle}[1][]{\savebox{\@brx}{\(\m@th{#1\langle}\)}%
  \mathopen{\copy\@brx\mkern2mu\kern-0.9\wd\@brx\usebox{\@brx}}}
\newcommand{\rrangle}[1][]{\savebox{\@brx}{\(\m@th{#1\rangle}\)}%
  \mathclose{\copy\@brx\mkern2mu\kern-0.9\wd\@brx\usebox{\@brx}}}
\makeatother

\newcommand{\yes}{\ding{51}}
\newcommand{\no}{\ding{55}}
\newcommand{\opt}{\matchlength{$^{\dag}$}{}}


\begin{document}

\title{Example-Based Reasoning about the Realizability of Polymorphic Programs}

\author{Niek Mulleners}
\orcid{0000-0002-7934-6834}
\affiliation{%
  \institution{Utrecht University}
  \city{Utrecht}
  \country{Netherlands}
}
\email{n.mulleners@uu.nl}

\author{Johan Jeuring}
\orcid{0000-0001-5645-7681}
\affiliation{%
  \institution{Utrecht University}
  \city{Utrecht}
  \country{Netherlands}
}
\email{j.t.jeuring@uu.nl}

\author{Bastiaan Heeren}
\orcid{0000-0001-6647-6130}
\affiliation{%
  \institution{Open University of the Netherlands}
  \city{Heerlen}
  \country{Netherlands}
}
\email{bastiaan.heeren@ou.nl}

\begin{abstract}
Parametricity states that polymorphic functions behave the same regardless of how they are instantiated. When developing polymorphic programs, Wadler's free theorems can serve as \emph{free specifications}, which can turn otherwise partial specifications into total ones, and can make otherwise realizable specifications \emph{unrealizable}. This is of particular interest to the field of program synthesis, where the unrealizability of a specification can be used to prune the search space. In this paper, we focus on the interaction between parametricity, input-output examples, and sketches. Unfortunately, free theorems introduce universally quantified functions that make automated reasoning difficult. Container morphisms provide an alternative representation for polymorphic functions that captures parametricity in a more manageable way. By using a translation to the container setting, we show how reasoning about the realizability of polymorphic programs with input-output examples can be automated.

\end{abstract}

\ccsdesc[500]{Software and its engineering~Programming by example}
\ccsdesc[500]{Theory of computation~Type theory}
\ccsdesc[300]{Theory of computation~Automated reasoning}

\keywords{parametricity, container functors, unrealizability, program synthesis,
example propagation}

\maketitle

\section{Introduction}

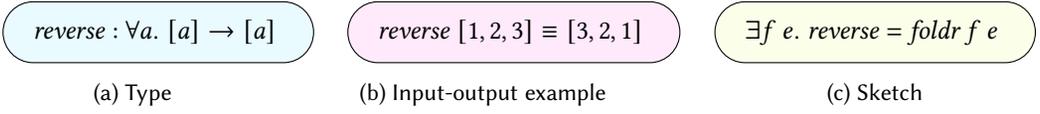
\begin{figure}
  \centering
  \begin{subfigure}[b]{0.25\textwidth}
    \centering
    \begin{tikzpicture}[auto,node distance=0mm]
      \node[spec,fill=type!25] {
        $\textit{reverse} : \forall a.\ [a] \rightarrow [a]$
      };
    \end{tikzpicture}
    \caption{Type}
    \label{fig:reverse type}
  \end{subfigure}
  \hfill
  \begin{subfigure}[b]{0.26\textwidth}
    \centering
    \begin{tikzpicture}[auto,node distance=0mm]
      \node[spec,fill=example!25] {
        $\textit{reverse}\ [1,2,3] \equiv [3,2,1]$
      };
    \end{tikzpicture}
    \caption{Input-output example}
    \label{fig:reverse example}
  \end{subfigure}
  \hfill
  \begin{subfigure}[b]{0.33\textwidth}
    \centering
    \begin{tikzpicture}[auto,node distance=0mm]
      \node[spec,fill=sketch!25] {
        $\exists f\ e.\ \textit{reverse} = \textit{foldr}\ f\ e$
      };
    \end{tikzpicture}
    \caption{Sketch}
    \label{fig:reverse sketch}
  \end{subfigure}
  \caption{A specification for the function \textit{reverse}.}
  \label{fig:reverse spec}
\end{figure}

The design of a program typically starts by specifying knowledge about the problem~\citep{felleisen_2018}. Often, this knowledge is in the form of \emph{types}, \emph{input-output examples}, and \emph{sketches} (incomplete programs containing holes). An example of each of these specifications for the function \textit{reverse} is shown in Figure~\ref{fig:reverse spec}. Types, examples, and sketches can help a programmer formulate a mental model of the problem at hand. Many modern languages also have native support for specifying types, examples, and sketches through type annotations, assertions, and holes. These specifications allow the programmer to express their intent to a programming environment and, in return, the environment could check whether the programmer's mental model is correct and catch possible mistakes early. To do so, the environment has to reason about the \emph{realizability} of a program as specified by the programmer, i.e.~whether a program adhering to this specification exists.

\begin{figure}
  \centering
  \begin{subfigure}[b]{0.25\textwidth}
    \centering
    \begin{tikzpicture}[auto]
      \node[spec,fill=type!25,dotted] {
        $f : \forall a\ b.\ a \rightarrow b$
      };
    \end{tikzpicture}
    \caption{An uninhabited type}
    \label{fig:unrealizable type}
  \end{subfigure}
  \hfill
  \begin{subfigure}[b]{0.3\textwidth}
    \centering
    \begin{tikzpicture}[auto]
      \node[spec,fill=example!25,dotted] {
        $f\ 1 \equiv 3 \;\land\; f\ 1 \equiv 5$
      };
    \end{tikzpicture}
    \caption{Contradicting examples}
    \label{fig:unrealizable examples}
  \end{subfigure}
  \hfill
  \begin{subfigure}[b]{0.33\textwidth}
    \centering
    \begin{tikzpicture}[auto]
      \node[spec,fill=type+example!25,dotted] {
        $f : \forall a.\ a \rightarrow a \;\land\; f\ 1 \equiv 3$
      };
    \end{tikzpicture}
    \caption{Parametricity at work}
    \label{fig:parametricity example}
  \end{subfigure}
  \caption{Specifications that are unrealizable.}
  \label{fig:unrealizable specs}
\end{figure}
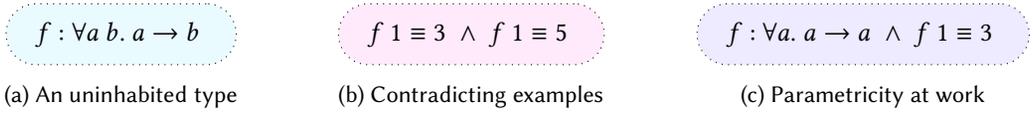

A type is realizable exactly if it is inhabited~\citep{urzyczyn_1997}. If a type is uninhabited, such as the type in Figure~\ref{fig:unrealizable type}, no program of that type exists. A set of input-output examples is realizable exactly if the examples do not contradict each other. If a set of input-output examples is contradictory, such as the examples in Figure~\ref{fig:unrealizable examples}, no program implementing those examples exists. Realizability of a type and realizability of a set of examples do not imply realizability of the combination! We are interested in exploring whether the intersections of program spaces as specified by types, examples, and sketches are empty. These intersections are visualized in Figure~\ref{fig:venn diagram}.

\def\typecircle{(150:12mm) circle (2cm)}
\def\polycircle{(150:12mm) circle (1.5cm)}
\def\examplecircle{(30:12mm) circle (2cm)}
\def\sketchcircle{(270:12mm) circle (2cm)}

\begin{figure}
  \centering
  \begin{subfigure}[t]{0.48\textwidth}
    \centering
    \begin{tikzpicture}[line width=0.2mm]
      \begin{scope}
        \fill[type!35] \typecircle;
        \fill[example!35] \examplecircle;
        \fill[sketch!35] \sketchcircle;
      \end{scope}
      \begin{scope}
        \clip \typecircle;
        \fill[type+example!35] \examplecircle;
      \end{scope}
      \begin{scope}
        \clip \examplecircle;
        \fill[example+sketch!35] \sketchcircle;
      \end{scope}
      \begin{scope}
        \clip \sketchcircle;
        \fill[sketch+type!35] \typecircle;
      \end{scope}
      \begin{scope}
        \clip \typecircle;
        \clip \examplecircle;
        \fill[type+example+sketch!35] \sketchcircle;
      \end{scope}
      \draw \typecircle node[above left=5mm] (x) {\Large $\mathcal{T}$};
      \draw \examplecircle node [above right=5mm] (y) {\Large $\mathcal{E}$};
      \draw \sketchcircle node [below=7mm] (z) {\Large $\mathcal{S}$};
    \end{tikzpicture}
    \caption{
    The intersections of programs of type $\tau$, implementing the set of examples $\varepsilon$, and refining the sketch $\varsigma$.
    }
    \label{fig:venn diagram}
  \end{subfigure}
  \hfill
  \begin{subfigure}[t]{0.48\textwidth}
    \centering
    \begin{tikzpicture}[auto,node distance=0mm]
      \node[spec,fill=type!25,minimum width=16mm] (b1) {$\mathcal{T}$};
      \node[right=of b1] (b12) {};
      \node[spec,right=of b12,fill=sketch!25,minimum width=16mm] (b2)
        {$\mathcal{S}$};
      \node[right=of b2] (b23) {};
      \node[spec,right=of b23,fill=example!25,minimum width=16mm] (b3)
        {$\mathcal{E}$};
      \node[below=11mm of b2] (b45) {};
      \node[spec,left=of b45,fill=sketch+type!25] (b4)
        {$\mathcal{T} \cap \mathcal{S}$};
      \node[spec,right=of b45,fill=example+sketch!25] (b5)
        {$\mathcal{S} \cap \mathcal{E}$};
      \node[spec,below=11mm of b45,fill=type+example+sketch!25] (b6)
        {$\mathcal{T} \cap \mathcal{S} \cap \mathcal{E}$};
      \node[above=3.5mm of b4.north] {\emph{check}};
      \node[above=3mm of b5.north] {\emph{propagate}};
      \node[above=3.5mm of b6.north] {\emph{translate}};
      \node[below right=of b6.south,align=left,xshift=0.4mm] {\emph{solve}};
      \node[below=9mm of b6] {};
      \begin{scope}[arr]
        \draw (b1.south) |- ($(b1)!0.55!(b4)$) -| (b4.north);
        \draw (b2.south) +(-1.5mm,0) |- ($(b2)!0.55!(b4)$) -| (b4.north);
        \draw (b2.south) +(1.5mm,0)  |- ($(b2)!0.55!(b5)$) -| (b5.north);
        \draw (b3.south) |- ($(b3)!0.55!(b5)$) -| (b5.north);
        \draw (b4.south) |- ($(b4)!0.55!(b6)$) -| (b6.north);
        \draw (b5.south) |- ($(b5)!0.55!(b6)$) -| (b6.north);
        \draw (b6.south) -- +(0,-6mm);
      \end{scope}
    \end{tikzpicture}
    \caption{The pipeline for computing the realizability of a type $\tau$, a set of examples $\varepsilon$, and a sketch $\varsigma$.
    }
    \label{fig:pipeline}
  \end{subfigure}
  \caption{$\mathcal{T}$ is the set of programs of type $\tau$. $\mathcal{E}$ is the set of programs implementing the examples $\varepsilon$. $\mathcal{S}$ is the set of programs refining the sketch $\varsigma$.}
  \label{fig:spec intersections}
\end{figure}
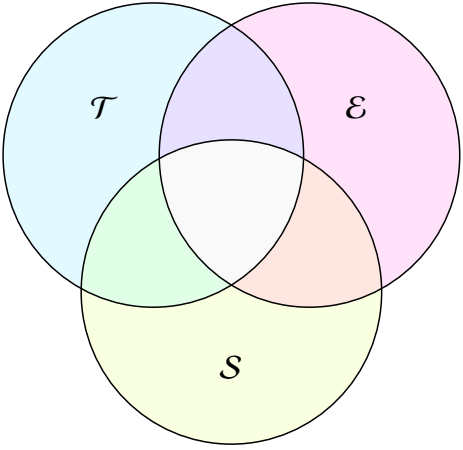
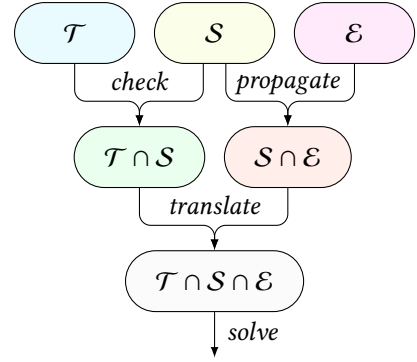

\paragraph*{Types and Examples}
The interaction between types and examples can be fairly subtle. Take for example the specification in Figure~\ref{fig:parametricity example}. Both the type and example are individually realizable, but \emph{parametricity} tells us that the only function with this type is the identity function~\citep{wadler_1989,reynolds_1983}, contradicting the input-output example. We are not aware of prior work that investigates the realizability of polymorphic programs with input-output examples. One reason for this could be that types and examples (as opposed to sketches) are typically assumed to be ground truth. This assumption does not hold when those types and examples are generated automatically. In particular, we will consider types and examples as propagated through a sketch.

\paragraph*{Types and Sketches}
Types and program sketching go hand in hand. Typed holes have recently been embraced by statically typed functional programming languages such as Haskell~\citep{haskell_holes, gissurarson_2018}, Agda~\citep{norell_2009}, and Hazel~\citep{omar_2019}. Holes allow partial programs to type check, while providing clear subgoals (hole types) for finishing those programs.

\paragraph*{Examples and Sketches}
Whereas examples are typically used as unit tests for complete programs, recent work has explored how and when incomplete programs (sketches) can be checked against input-output examples, resulting in new input-output examples on the holes~\citep{omar_2019,lubin_2020,feser_2015,mulleners_2023}. We refer to this process as \emph{example propagation}.

\paragraph*{Types, Examples, and Sketches}
When inferring types and input-output examples for the holes of a sketch, reasoning about the realizability of those types and examples becomes a worthy avenue, since the realizability of a sketch relies on the realizability of its holes. For example, assume a programmer specifies \textit{reverse} as in Figure~\ref{fig:reverse spec}, but instead incorrectly writes the sketch $\textit{reverse} = \textit{map}\ f$. To finish the program, there has to exist a function $f$, i.e.~the realizability of \textit{reverse} with this sketch implies the realizability of $f$. The sketch can be evaluated against the input-output example by inlining the definitions of \textit{map} and equality on lists:
\[
  \textit{map}\ f\ [1,2,3] \equiv [3,2,1]
  \implies
  [f\ 1,f\ 2,f\ 3] \equiv [3,2,1]
  \implies
  f\ 1 \equiv 3 \land f\ 2 \equiv 2 \land f\ 3 \equiv 1
\]
From the type of \textit{reverse}, the type of $f$ is inferred to be $a \rightarrow a$, for some fixed $a$. As we saw before, the only function of this type is the identity function, which contradicts the input-output examples. The unrealizability of $f$ allows us to correctly conclude that the sketch $\textit{reverse} = \textit{map}\ f$ is unrealizable.

\paragraph*{Parametricity}
It should be possible to prove the unrealizability of \textit{reverse} in terms of \textit{map} directly using Reynolds' abstraction theorem~\citep{reynolds_1983} or Wadler's free theorems~\citep{wadler_1989}. However, such proofs do not generalize easily. More importantly, we are not just interested in proving the realizability of arbitrary specifications. Rather, we want a decision procedure that decidably computes whether a specification is realizable or not. Intuitively, a parametrically polymorphic function cannot \emph{inspect} or \emph{produce} polymorphic elements, only ``pass them around". This intuition is perfectly captured by \emph{container functors} and \emph{container morphisms}~\citep{abbot_2005}, which provide normal forms for polymorphic datatypes and polymorphic functions respectively. \newline

In this paper, we propose a general procedure for computing the realizability of polymorphic programs: first, \emph{type checking} and \emph{example propagation} are used to infer types and input-output examples for the holes of a sketch; then, the resulting constraints are \emph{translated} to the container setting; lastly, the translated constraints are \emph{solved} using an SMT solver. This procedure is visualized in Figure~\ref{fig:pipeline}. More specifically, our contributions are to show how to
\begin{itemize}
  \item compute the realizability of polymorphic types with monomorphic input-output examples (Section~\ref{sec:formalization});
  \item use example propagation to extend this reasoning to sketching with higher-order combinators such as \textit{map} (Section~\ref{sec:map}); 
  \item reason about input-output traces to support sketching with recursion schemes such as \textit{foldr} (Section~\ref{sec:traces});
  \item generalize the notion of trace completeness to take parametricity into account (Section~\ref{sec:foldr}).
\end{itemize}

\section{Background: Containers} \label{sec:container functors}

Any list can be uniquely represented by its length $n$ and a function $f$ assigning a value to every index in $\{0,1,\dots,n-1\}$. For example, the list $[A,B,C]$ can be represented by $n = 3$, $f\ 0 = A$, $f\ 1 = B$, and $f\ 2 = C$. We say that a list of length $n$ is a \emph{container} with \emph{shape} $n$ and \emph{positions} $\{0,1,\dots,n-1\}$. Many datatypes can similarly be represented in terms of their shape and positions.

Containers~\cite{abbot_2003} present a generic way of modeling datatypes that \emph{contain} elements, making the distinction between the structure and the content of a datatype explicit. A container $S \triangleright P$ (sometimes written $(s : S) \triangleright P_s$) is defined by a type of shapes $S$, representing the structure of the datatype, and for every shape $s$ a type of positions $P_s$, describing where the elements are located.
\begin{definition} \label{def:list container}
  The list container is defined as $\Nat \triangleright \Fin$, where $\Fin\ n = \{0,1,\dots,n-1\}$, i.e.~the type of natural numbers smaller than $n$. A list $[x_0,\dots,x_{n-1}]$ is represented using the list container as a pair $(n, \lambda i.\ x_i)$.
\end{definition}
\noindent
Functors that can be defined as containers are called \emph{container functors}. A container functor $F$ is isomorphic to the \emph{extension} of some container $S \triangleright P$, defined as $\llbracket S \triangleright P \rrbracket\ a = \Sigma_{(s:S)}.\ (P_s \rightarrow a)$. In other words, if $F$ is a container functor corresponding to the container $S \triangleright P$, values of type $F\,a$ can be represented by a pair $(s, p)$, where $s$ is a shape of type $S$ and $p$ is a function assigning an element of type $a$ to every position $P_s$.

\subsection{Constructing Containers}

The simplest container is the identity container, having only a single shape and a single position.
\begin{definition}
  The identity container is defined as $\M1 \triangleright \K\ \M1$, where $\M1$ is the unit type with a single element $\star$ and $\K$ is the constant function, i.e.~$\K\ x\ y = x$. A value $x : a$ is represented using the identity container as a pair $(\star, \K\ x)$.
\end{definition}
\noindent
Containers can be combined to create more complex containers. A pair of lists, for example, has two natural numbers $n_1$ and $n_2$ as its shape (the lengths of the lists) and exactly $n_1 + n_2$ positions where elements are stored.
\begin{definition} \label{def:product container}
  The product of two containers $S \triangleright P$ and $T \triangleright
  Q$ is defined as $((s, t) : S \times T) \triangleright (P_s + Q_t)$. Given
  that the container representation of $x$ and $y$ are $(s, p)$ and $(t, q)$
  respectively, the pair $(x, y)$ is represented as $((s, t), p \oplus q)$,
  where $(p \oplus q)\ (\textbf{\upshape inl}\ z) = p\ z$ and $(p \oplus q)\
  (\textbf{\upshape inr}\ z) = q\ z$.
\end{definition}
\noindent
Using combinators like these, containers can be used to construct all \emph{strictly positive types}, which can intuitively be understood as all types that have a tree-like structure. Additionally, the translation between values of inductively defined strictly positive types and their container representation can be automated~\cite{abbot_2005}. As a convention, we use $\shape{\color{gray} \bullet}$ to denote the translated shape component and $\pos{\color{gray} \bullet}$ to denote the translated position component. For example, the functor $F$ corresponds to the container $\shape{F} \triangleright \pos{F}$ and a value $x$ of type $F\ a$ is translated to a pair $(\shape{x}, \pos{x})$.

\subsection{Container Morphisms}

Given a list $[x_0,x_1,\dots,x_{n-1}]$, the reverse of that list is $[x_{n-1},x_{n-2},\dots,x_0]$. Using the container representation of lists, the reverse of the list $(n, \lambda i.\ x_i)$ is the list $(n, \lambda i.\ x_{n - i - 1})$. Note how we can describe list reversal by specifying, for each index $i$ in the output list, where the element at that index comes from in the input list: the index $n - i - 1$. We can use arrows to visualize where each element comes from:
\[
\begin{tikzpicture}[]
  \matrix [column sep=-1.5mm]
  {
    \node{$\textit{reverse}\ [$}; &
    \node[gray](a) {A}; &
    \node[yshift=-1mm]{,}; & \node[gray](b) {B}; &
    \node[yshift=-1mm]{,}; & \node[gray](c) {C}; &
    \node(m) {$] \equiv [$}; &
    \node[gray](d) {C}; &
    \node[yshift=-1mm]{,}; & \node[gray](e) {B}; &
    \node[yshift=-1mm]{,}; & \node[gray](f) {A}; &
    \node{]}; \\
  };
  \draw [stealth-] (a.north) |- ([yshift=5.5mm]m.north) -| (f.north);
  \draw [stealth-] (b.north) |- ([yshift=3.5mm]m.north) -| (e.north);
  \draw [stealth-] (c.north) |- ([yshift=1.5mm]m.north) -| (d.north);
\end{tikzpicture}
\]
Similarly, the tail of the list $(n, \lambda i.\ x_i)$ is the list $(n - 1, \lambda i.\ x_{i + 1})$:
\[
\begin{tikzpicture}[]
  \matrix [column sep=-1.5mm]
  {
    \node{$\textit{tail}\ [$}; &
    \node[gray](a) {A}; &
    \node[yshift=-1mm]{,}; & \node[gray](b) {B}; &
    \node[yshift=-1mm]{,}; & \node[gray](c) {C}; &
    \node(m) {$] \equiv [$}; &
    \node[gray](d) {B}; &
    \node[yshift=-1mm]{,}; & \node[gray](e) {C}; &
    \node{]}; \\
  };
  \draw [stealth-] (b.north) |- ([yshift=1.5mm]m.north) -| (d.north);
  \draw [stealth-] (c.south) |- ([yshift=-1.5mm]m.south) -| (e.south);
\end{tikzpicture}
\]
This idea of defining a function in terms of how the shape changes and where each element in the output comes from is formalized using \emph{container morphisms}. A container morphism between the containers $S \triangleright P$ and $T \triangleright Q$ is a pair of functions $(u, g)$, where $u : S \rightarrow T$ maps input shapes to output shapes and $g : \Pi_{s:S}(Q_{(u\ s)} \rightarrow P_s)$ maps output positions to input positions~\citep{abbot_2003}. Notice again the explicit separation between structure and content. The \emph{extension} of a container morphism $(u, g)$ (written $\llangle u, g \rrangle$) is a function mapping $(s, p)$ to $(u\ s, p \circ g_s)$. We can define \textit{reverse} and \textit{tail} as the extension of a container morphism as follows:
\[
  \textit{reverse} = \llangle (\lambda n.\ n), (\lambda n\ i.\ n - i - 1)
  \rrangle
  \qquad\qquad
  \textit{tail} = \llangle (\lambda n.\ n - 1), (\lambda n\ i.\ i + 1) \rrangle
\]
A container morphism represents a natural transformation between container functors. Moreover, every natural transformation between container functors can be defined as the extension of a container morphism~\citep{abbot_2005}.

\subsection{Small Containers}

\emph{Small containers} (also known as \emph{finitary containers}) are containers whose shapes have a decidable equality and whose positions are finite sets~\citep{prince_2008}. The list container is an example of a small container. Even though a list may contain any number of elements, for any given shape it has a finite number of positions. All the containers in this paper are small, enabling us to enumerate over their positions, which is crucial in decidably computing the realizability of a program.

\section{Realizability of Input-Output Examples with Polymorphic
Types}\label{sec:formalization}

To reason about the realizability of an input-output example $f\ x \equiv y$, where $f$ is a natural transformation, we assume that there exists a container morphism $(\shape{f},\pos{f})$ satisfying this example. To do so, we have to translate the input $x$ and the output $y$ to the container setting. If we can show that such a container morphism $(\shape{f},\pos{f})$ is not realizable, then neither is the natural transformation $f$.

\newcommand{\invar}{x}
\newcommand{\outvar}{y}
\newcommand{\inpos}{p}
\newcommand{\outpos}{q}

\subsection{Translating a Single Example}

An input-output example for $f : \forall a.\ F\,a \rightarrow G\,a$ is defined by a triple $(\tau, \invar, \outvar)$, where $\tau$ is a monomorphic type, $\invar$ is an input of type $F\,\tau$, and $\outvar$ is an output of type $G\, \tau$, such that $f\ \invar \equiv \outvar$. Given that $F$ and $G$ are container functors isomorphic to $\llbracket \shape{F} \triangleright \pos{F} \rrbracket$ and $\llbracket \shape{G} \triangleright \pos{G} \rrbracket$ respectively, we can translate $\invar$ and $\outvar$ to the container setting. The input $\invar$ is translated to a pair $(\shape{\invar}, \pos{\invar})$, where $\shape{\invar} : \shape{F}$ and $\pos{\invar} : \pos{F}_{\shape{\invar}} \rightarrow \tau$ and the output $\outvar$ is translated to a pair $(\shape{\outvar}, \pos{\outvar})$, where $\shape{\outvar} : \shape{G}$ and $\pos{\outvar} : \pos{G}_{\shape{\outvar}} \rightarrow \tau$. Since $f$ is a natural transformation between $F$ and $G$, it corresponds to a container morphism $(\shape{f}, \pos{f})$ between $\shape{F} \triangleright \pos{F}$ and $\shape{G} \triangleright \pos{G}$. There exists a function $f$ such that $f\ \invar \equiv y$ exactly if there exists a container morphism $(\shape{f}, \pos{f})$ such that $\llangle \shape{f}, \pos{f} \rrangle\ (\shape{\invar}, \pos{\invar}) \equiv (\shape{\outvar}, \pos{\outvar})$:
\begin{center}
\begin{tikzpicture}[auto,node distance=0mm]
  \node[spec,fill=type!25] (b1) {
    $\forall a.\ f : F\,a \rightarrow G\,a$
  };
  \node[left=of b1] {$\exists f.$};
  \node[right=of b1] (b12) {$\land$};
  \node[spec,right=of b12,fill=example!25] (b2) {
    $f\ \invar \equiv \outvar$
  };
  \node[spec,below=8mm of b12,fill=type+example!25] (b3) {
    $\llangle \shape{f}, \pos{f} \rrangle\ (\shape{\invar}, \pos{\invar})
    \equiv (\shape{\outvar}, \pos{\outvar})$
  };
  \node[left=of b3] {$\exists (\shape{f},\pos{f}).$};
  \begin{scope}[arr]
    \draw (b1.south) |- ($(b1)!0.45!(b3)$) -| (b3.north);
    \draw (b2.south) |- ($(b2)!0.45!(b3)$) -| (b3.north);
  \end{scope}
\end{tikzpicture}
\end{center}
To solve the equation $\llangle \shape{f}, \pos{f} \rrangle\ (\shape{\invar}, \pos{\invar}) \equiv (\shape{\outvar}, \pos{\outvar})$, we first rewrite it by applying the container morphism extension, resulting in $(\shape{f}\ \shape{\invar}, \pos{\invar} \circ \pos{f}) \equiv (\shape{\outvar}, \pos{\outvar})$.\footnote{For the sake of readability, we leave the dependent argument to $\pos{f}$ implicit, as it can be inferred from the type of its argument.} Using equivalence on containers, as well as function extensionality, we separate the \textit{shape} constraints on $\shape{f}$ from the \textit{position} constraints on $\pos{f}$:
\[
  \underbrace{\shape{f}\ \shape{\invar} \equiv \shape{\outvar}}_{\textit{shape}}
  \;\land\;
  \underbrace{
  \forall \outpos.\ \pos{\invar}\ (\pos{f}\ \outpos) \equiv
  \pos{\outvar}\ \outpos}
  _{\textit{position}}
\]
Next, we make the intermediate argument returned by $\pos{f}$ explicit by introducing an existential quantification:
\begin{equation}\label{eq:morphism realizability}
  \underbrace{\shape{f}\ \shape{\invar} \equiv \shape{\outvar}}_
  {\text{\ref{shape realizability}}}
  \;\land\;
  \forall \outpos.\ \exists \inpos.\ \underbrace{\pos{f}\ \outpos \equiv
  \inpos}_{\text{\ref{position realizability}}}
  \;\land\;
  \underbrace{\pos{\invar}\ \inpos \equiv \pos{\outvar}\ \outpos}_
  {\text{\ref{example consistency}}}
\end{equation}
The resulting equation consists of three components:
\begin{enumerate}[label=(\alph*)]
  \item\label{shape realizability} An input-output example for $\shape{f}$, describing the realizability of the shape morphism.
  \item\label{position realizability} A set of input-output examples for $\pos{f}$, describing the realizability of the position morphism.
  \item\label{example consistency} A consistency check with respect to parametricity, ensuring that each element in the output also occurs in the input. Note how $\forall \outpos.\ \exists \inpos.\ \pos{\invar}\ \inpos \equiv \pos{\outvar}\ \outpos$ can be rewritten as $\textit{codomain}(\pos{\outvar}) \subseteq \textit{codomain}(\pos{\invar})$.
\end{enumerate}
Similar to how the consistency of input-output examples with respect to a sketch can be computed using example propagation, resulting in input-output examples for the subcomponents of the program (the holes), our translation to the container setting checks the consistency of input-output examples with respect to a polymorphic type, resulting in input-output examples for the abstract subcomponents of the program (the shape and position morphisms).

To figure out if $f$ is realizable, we have to solve the satisfiability of Equation~\ref{eq:morphism realizability}, but this is complicated by the universal and existential quantifications. If we assume, however, that $\shape{F} \triangleright \pos{F}$ and $\shape{G} \triangleright \pos{G}$ are \emph{small containers}, we can get rid of any quantifications by enumerating over the positions:
\begin{equation}\label{eq:small morphism realizability}
  \shape{f}\ \shape{\invar} \equiv \shape{\outvar} \;\land\; \bigwedge_\outpos
  \bigvee_\inpos\; \pos{f}\ \outpos \equiv \inpos \;\land\;
  \pos{\invar}\ \inpos \equiv \pos{\outvar}\ \outpos
\end{equation}
When we know $\pos{\invar}$ and $\pos{\outvar}$, the satisfiability of this formula is trivially determined by an SMT solver.

\subsection{Example: Reverse as Map}

To see how example consistency can be used to reason about the realizability of a program, we will turn our attention once more to the example of \textit{reverse} in terms of \textit{map}:
\begin{center}
\begin{tikzpicture}[auto,node distance=0mm]
  \node[spec,fill=type!25] (b1) {
    $\forall a.\ \textit{reverse} : [a] \rightarrow [a]$
  };
  \node[right=of b1] (b12) {$\land$};
  \node[spec,right=of b12,fill=sketch!25] (b2) {
    $\exists f.\ \textit{reverse} = \textit{map}\ f$
  };
  \node[right=of b2] (b23) {$\land$};
  \node[spec,right=of b23,fill=example!25] (b3) {
    $\textit{reverse}\ [A,B,C] \equiv [C,B,A]$
  };
\end{tikzpicture}
\end{center}
The first step in checking the realizability of \textit{reverse} is to propagate the type and input-output example through the sketch. Type checking gives $f : a \rightarrow a$, for a fixed $a$. To propagate the example, we evaluate the sketch applied to the input $[A,B,C]$ as far as possible, by inlining the definition of \textit{map}:
\[
  \textit{map}\ f\ [A,B,C]
  \quad\leadsto\quad
  [f\ A,f\ B,f\ C]
\]
Then, we simplify the resulting equation using equality on lists:
\[
  [f\ A,f\ B,f\ C] \equiv [C, B, A]
  \quad\implies\quad
  {f\ A \equiv C \color{gray} \;\land\; f\ B \equiv B \;\land\; f\ C \equiv A}
\]
We will focus on the first conjunct (highlighted): intuitively, $f\ A \equiv C$ is inconsistent, because the $C$ in the output does not occur in the input of $f$. This implies that $f$ is unrealizable, which we can prove formally by translating the example $f\ A \equiv C$ to the container setting. We interpret $f$ as a natural transformation $(\shape{f},\pos{f})$ between the identity functor and translate the values $A$ and $C$ accordingly:
\begin{center}
\begin{tikzpicture}[auto,node distance=0mm]
  \node[spec,fill=type!25] (b1) {
    $\forall a.\ f : a \rightarrow a$
  };
  \node[left=of b1] {$\exists f.$};
  \node[right=of b1] (b12) {$\land$};
  \node[spec,right=of b12,fill=example!25] (b2) {
    $f\ A \equiv C$
  };
  \node[spec,below=8mm of b12,fill=type+example!25] (b3) {
    $\llangle \shape{f}, \pos{f} \rrangle\
    (\star, \K\ A)
    \equiv
    (\star, \K\ C)
    $
  };
  \node[left=of b3] {$\exists (\shape{f},\pos{f}).$};
  \begin{scope}[arr]
    \draw (b1.south) |- ($(b1)!0.45!(b3)$) -| (b3.north);
    \draw (b2.south) |- ($(b2)!0.45!(b3)$) -| (b3.north);
  \end{scope}
\end{tikzpicture}
\end{center}
We write out the resulting constraint on $(\shape{f},\pos{f})$ according to Equation~\ref{eq:morphism realizability} and simplify:
\[
  \underbrace{\shape{f}\ \star \equiv \star}_{\text{\ref{shape realizability}}}
  \;\land\;
  \underbrace{\pos{f}\ \star \equiv \star}_{\text{\ref{position realizability}}}
  \;\land\;
  \underbrace{\phantom{f}\kern-1.7mm A \equiv C
  }_{\text{\ref{example consistency}}}
\]
The constraints on $\shape{f}$ and $\pos{f}$ are trivially satisfied, but the consistency check~\ref{example consistency} fails: the codomain of $\K\ C$ is not a subset of the codomain of $\K\ A$, i.e.~ $\{C\} \not\subseteq \{A\}$. We conclude that $f$, and, by extension, \textit{reverse} as a \textit{map}, is unrealizable.

\subsection{Translating Multiple Examples}

The only way to show that a single input-output example is unrealizable is by showing that it fails the consistency check~\ref{example consistency}. However, when considering multiple input-output examples, their realizability additionally relies on the realizability of their respective shape morphisms~\ref{shape realizability} and position morphisms~\ref{position realizability}.

A polymorphic function $f : \forall a.\ [a] \rightarrow [a]$ cannot \emph{inspect} the elements in the input list and can therefore not differentiate between different input lists that have the same length (i.e.~the same \emph{shape}). When a set of input-output examples for $f$ requires inspecting the input elements, this will result in either a shape conflict on $\shape{f}$ or a position conflict on $\pos{f}$. For example, assume that $f$ is supposed to sort the input list and remove any duplicates. This requires inspecting the elements in the list, so we expect to find a contradiction:
\begin{itemize}
  \item The examples $f\ [A, A] \equiv [A] \land f\ [A,B] \equiv [A,B]$ imply
  the contradictory shape constraint $\shape{f}\ 2 \equiv 1 \land \shape{f}\ 2
  \equiv 2$, leading to a \emph{shape conflict}.
  \item The examples $f\ [A,B] \equiv [A,B] \land f\ [B,A] \equiv [A,B]$ imply
  the contradictory position constraint $\pos{f}\ 0 \equiv 0 \land \pos{f}\ 0
  \equiv 1$, leading to a \emph{position conflict}.
\end{itemize}

\subsection{Sketching with Map}\label{sec:map}

\begin{figure}
  \centering
  \begin{tikzpicture}[auto,node distance=0mm]
    \node[spec,fill=type!25] (b1) {
      $\forall a.\ p : [F\,a] \rightarrow [G\,a]$
    };
    \node[left=of b1] {$\exists p.$};
    \node[right=of b1] (b12) {$\land$};
    \node[spec,right=of b12,fill=sketch!25] (b2) {
      $\exists f.\ p = \textit{map}\ f$
    };
    \node[right=of b2] (b23) {$\land$};
    \node[spec,right=of b23,fill=example!25] (b3) {
      $p\ [\invar_0 \cdots \invar_{n-1}] \equiv [\outvar_0 \cdots \outvar_{n-1}]$
    };
    \node[below=9mm of b2] (b45) {$\land$};
    \node[spec,left=of b45,fill=sketch+type!25] (b4) {
      $\forall a.\ f : F\,a \rightarrow G\,a$
    };
    \node[left=of b4] {$\exists f.$};
    \node[spec,right=of b45,fill=example+sketch!25,stacked] (b5) {
      $f\ \invar_i \equiv \outvar_i$
    };
    \node[right=1.5mm of b5,yshift=-1.5mm] {$_{0 \leq i < n}$};
    \node[spec,below=9mm of b45,fill=type+example+sketch!25,stacked] (b6) {
      $\llangle \shape{f}, \pos{f} \rrangle\ (\shape{\invar}_i, \pos{\invar}_i)
      \equiv (\shape{\outvar}_i, \pos{\outvar}_i)$
    };
    \node[left=of b6] {$\exists (\shape{f},\pos{f}).$};
    \node[right=1.5mm of b6,yshift=-1.5mm] {$_{0 \leq i < n}$};
    \begin{scope}[arr]
      \draw (b1.south) |- ($(b1)!0.5!(b4)$) -| (b4.north);
      \draw (b2.south) +(-3mm,0) |- ($(b2)!0.5!(b4)$) -| (b4.north);
      \draw (b2.south) +(3mm,0)  |- ($(b2)!0.5!(b5)$) -| (b5.north);
      \draw (b3.south) |- ($(b3)!0.5!(b5)$) -| (b5.north);
      \draw (b4.south) |- ($(b4)!0.5!(b6)$) -| (b6.north);
      \draw (b5.south) +(0,-0.6mm) |- ($(b5)!0.5!(b6)$) -| (b6.north);
    \end{scope}
  \end{tikzpicture}
  \caption{The pipeline for computing the realizability of a polymorphic
  program defined as a \textit{map} with a single input-output example.}
  \label{fig:map pipeline}
\end{figure}
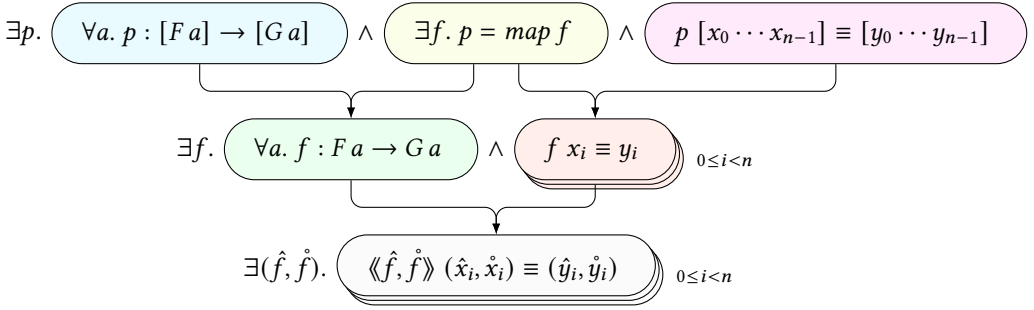

Now that we have seen how to reason about the realizability of sets of input-output examples, it is straightforward to extend this reasoning to any program $p$ specified with the sketch $\exists f.\ p = \textit{map}\ f$, by recognizing that an input-output example for $p$ represents a set of input-output examples for $f$. To be precise, an input-output example for $p$ with an input list of length $n$ corresponds to exactly $n$ input-output examples for $f$. The complete pipeline for programs defined as a \textit{map} is shown in Figure~\ref{fig:map pipeline}. Note how each step in the pipeline changes the focus: first the whole program $p$; then the hole $f$; and finally the container morphism $(\shape{f},\pos{f})$.

\paragraph*{Soundness}
Realizability reasoning is sound if we cannot draw false conclusions from its results. This is the case exactly if the initial equation in the pipeline is equivalent to the final equation. That way, a solution to the final equation implies that a correct program exists; and a contradiction implies that no such program exists. We argue that reasoning about the realizability of a program as a \textit{map} is sound, as all the steps shown in Figure~\ref{fig:map pipeline} are valid rewritings.

\paragraph*{Completeness}
Realizability reasoning is complete exactly if it is a decision procedure, i.e.~it always returns either a solution or a contradiction. This is true if the final equation in the pipeline is in a decidable logic. Reasoning about the realizability of a program as a \textit{map} is complete as long as $F$ and $G$ are small container functors. That way, the final equation in Figure~\ref{fig:map pipeline} is in a quantifier free logic, where each component is an instantiation of Equation~\ref{eq:small morphism realizability}.

\section{Reasoning About Recursion Schemes}\label{sec:traces}

One of the most ubiquitous functions in functional programming is \textit{foldr}, a recursion scheme that embodies structural recursion over inductively defined datastructures. \citet{gibbons_2001} ask the question ``When is a function a fold?" and show how to prove whether a function is indeed a fold. Intuitively speaking, and restricting ourselves to folds over lists, we say that a function $h$ is a fold if, for any element $x$ and any list \textit{xs}, the result of $h\ (x:\textit{xs})$ is uniquely determined by $x$ and $h\ \textit{xs}$. For example, the function \textit{reverse} is a fold, since
\[
  \textit{reverse}\ (x:\textit{xs}) =
  \textit{reverse}\ \textit{xs} \doubleplus [x].
\]
The function $\textit{tail} : \forall a.\ [a] \rightarrow [a]$ (taken from \citep{gibbons_2001}), however, given by the equations
\[\arraycolsep=1.4pt
\begin{array}{lcl}
  \textit{tail}\ [] &=& []\\
  \textit{tail}\ (x : \textit{xs}) &=& \textit{xs}\\
\end{array}
\]
is not a fold, because $\textit{tail\ \textit{xs}}$ throws away the first element of \textit{xs}, which is needed in the tail of $x:\textit{xs}$. In Haskell terms, \textit{tail} is not a fold because there exist no $f$ and $e$ such that $\textit{tail} = \textit{foldr}\ f\ e$.

\subsection{Realizability of Tail}

\citet{hofmann_2010b} shows how to detect whether a set of input-output examples specifies a fold. Similar results can be achieved using example propagation~\citep{feser_2015,lubin_2020}. In this section, we will use a combination of example propagation and parametric reasoning to show that \textit{tail} is \emph{not} a fold using a minimal number of input-output examples.

To prove automatically that \textit{tail} is not a fold, we will prove that $\textit{tail} : \forall a.\ [a] \rightarrow [a]$ with the sketch $\exists f.\ \textit{tail} = \textit{foldr}\ f\ []$ is unrealizable:\footnote{Note that this defines $\textit{tail}\ [] = []$.}
\begin{center}
\begin{tikzpicture}[auto,node distance=0mm]
  \node[spec,fill=type!25] (b1) {
    $\forall a.\ \textit{tail} : [a] \rightarrow [a]$
  };
  \node[right=of b1] (b12) {$\land$};
  \node[spec,right=of b12,fill=sketch!25] (b2) {
    $\exists f.\ \textit{tail} = \textit{foldr}\ f\ []$
  };
  \node[right=of b2] (b23) {$\land$};
  \node[spec,right=of b23,fill=example!25] (b3) {
    $\matchlength{\kern-0.85mm\dots}{}$
  };
\end{tikzpicture}
\end{center}
\noindent
Whether and how we can prove the unrealizability of this specification depends on the exact input-output examples. For example, take the following input-output examples for \textit{tail}:
\begin{center}
\begin{tikzpicture}[auto]
  \node[spec,fill=example!25] {
    $\textit{tail}\ [A,B,C] \equiv [B, C] \;\land\; \textit{tail}\ [D,E] \equiv [E]$
  };
\end{tikzpicture}
\end{center}
Since \textit{foldr} uses structural recursion, the call $\textit{tail}\ [A,B,C]$ relies on recursive calls to $[B,C]$, $[C]$, and $[]$. Because we cannot inspect the elements, we cannot distinguish the call to $[B,C]$ from the call to $[D,E]$. Since the call to $[D,E]$ returns $[E]$, throwing away the first element, the element $B$ is also thrown away in the call to $[B,C]$. This implies that the result of tail $[A,B,C]$ cannot contain the value $B$, contradicting our example.

Formally, the proof starts with propagating the type and input-output examples through the sketch. Given the types of \textit{tail} and \textit{foldr}, the type of $f$ is inferred to be $a \times [a] \rightarrow [a]$, for some fixed $a$. To propagate the example $\textit{tail}\ [A,B,C] \equiv [B,C]$, we evaluate the sketch applied to the input $[A,B,C]$, by inlining the definition of \textit{foldr}:
\[
  \textit{foldr}\ f\ []\ [A,B,C]
  \quad\leadsto\quad
  f\ (A,f\ (B,f\ (C,[])))
\]
Unfortunately, the resulting equation $f\ (A,f\ (B,f\ (C,[]))) \equiv [B,C]$ is \emph{not} an input-output example for $f$, but rather an input-output \emph{trace}, specifying the trace of calls made to $f$ resulting in the output $[B,C]$. We will make the intermediate results returned by $f$ explicit by introducing two existentially quantified variables, revealing that an input-output trace represents a set of related input-output examples:
\begin{center}
\begin{tikzpicture}[auto]
  \node[spec,fill=example+sketch!25] {
    $\exists x\ y.\ f\ (A,x) \equiv [B,C] \land
    f\ (B,y) \equiv x \land f\ (C,[]) \equiv y$
  };
\end{tikzpicture}
\end{center}
To translate these examples, we interpret $f$ as a natural transformation $(\shape{f},\pos{f})$ between the non-empty list container (i.e.~the product of the identity container and the list container) and the list container. For each existentially quantified variable $x$, we introduce two existentially quantified variables $\shape{x}$ and $\pos{x}$:
\newcommand{\y}{\matchlength{$y$}{x}}
\[\arraycolsep=1.4pt
  \begin{array}{rll}
  \exists (\shape{x},\pos{x})\ (\shape{\y},\pos{\y}). &
  \llangle \shape{f},\pos{f}
  \rrangle\ ((\star,\shape{x}), \K\ A \oplus \pos{x})
  &\equiv (2, \{0 \mapsto B, 1 \mapsto C \})\\
  \land& 
  \llangle \shape{f},\pos{f}
  \rrangle\ ((\star,\shape{\y}), \K\ \matchlength{$B$}{A} \oplus \pos{\y})
  &\equiv (\shape{x},\pos{x})\\
  \land&
  \llangle \shape{f},\pos{f} \rrangle\ 
  ((\star,\matchlength{0}{\shape{x}}), \K\ C) &\equiv (\shape{\y},\pos{\y})\\
  \end{array}
\]
We first focus on the position component of $f\ (A,x) \equiv [B,C]$. After simplifying:
\[
  (\K\ A \oplus \pos{x})(\pos{f}\ 0) \equiv B
  \land (\K\ A \oplus \pos{x})(\pos{f}\ 1) \equiv C
\]
The constant function $\K\ A$ cannot return $B$ or $C$, implying that both $B$ and $C$ are in the codomain of $\pos{x}$:
\begin{equation}\label{eq:tail codomain}
  \{B, C\} \subseteq \textit{codomain}(\pos{x})
\end{equation}
In other words, both $B$ and $C$ are returned by the recursive call $\textit{tail}\ [B,C]$. Given the polymorphic type of \textit{tail}, this should conflict the constraint $\textit{tail}\ [D,E] \equiv [E]$. To prove this, we make the input-output trace incurred by $\textit{tail}\ [D,E] \equiv [E]$ explicit:
\begin{center}
\begin{tikzpicture}[auto]
  \node[spec,fill=example+sketch!25] {
    $\exists z.\ f\ (D,z) \equiv [E] \land f\ (E,[]) \equiv []$
  };
\end{tikzpicture}
\end{center}
Next, we will match up the shape components of the different calls to $f$: the input shapes of $f\ (C, [])$ and $f\ (E, [])$ are equal, so we can unify their output shapes $\shape{y}$ and $\shape{z}$; in doing so, the input shapes of $f\ (B, y)$ and $f\ (D, z)$ become equal, showing that $\shape{x}$ is equal to 1:
\[\arraycolsep=1.4pt
  \left.
  \begin{array}{l}
    \left.
    \kern-0.4mm
    \begin{array}{ll}
      f\ (\matchlength{$C$}{D}, []) \equiv \matchlength{$y$}{[E]}
        & \;\implies\;
        \shape{f}\ (\star, \matchlength{$0$}{y}) \equiv \shape{y} \\
      f\ (\matchlength{$E$}{D}, []) \equiv z
        & \;\implies\;
        \shape{f}\ (\star, \matchlength{$0$}{y}) \equiv \shape{z} \\
    \end{array}
    \;\; \right\} \;\; \shape{y} \equiv \shape {z} \\
    \begin{array}{ll}
      f\ (\matchlength{$B$}{D}, \matchlength{$\,y$}{[]}) \equiv x
        & \;\implies\;
        \shape{f}\ (\star, \matchlength{$\shape{y}$}{y}) \equiv \shape{x} \\
      f\ (D, \matchlength{$\,z$}{[]}) \equiv \matchlength{$[E]$}{[E]}
        & \;\implies\;
        \shape{f}\ (\star, \matchlength{$\shape{z}$}{y}) \equiv 1
    \end{array}
  \end{array}
  \quad \right\} \quad \shape{x} \equiv 1
\]
This implies that the domain of $\pos{x} : \textit{Fin}\ \shape{x} \rightarrow \tau$ is equal to $\textit{Fin}\ 1$. In other words, we used the shape component of $\textit{tail}\ [D,E] \equiv [E]$ to show that the recursive call to $\textit{tail}\ [B,C]$ returns a list with a single element. Naturally, this list cannot contain both $B$ and $C$, contradicting \eqref{eq:tail codomain} and thereby proving that \textit{tail} is not a fold. Finding this contradiction by hand requires some effort, but this is trivial for an SMT solver.

\paragraph*{Generalizing Beyond Tail}\label{sec:generalize}

Adding a constant, monomorphic input to a set of examples does not affect realizability. Hence, our proof readily extends to functions that generalize over \textit{tail} using an additional argument:
\begin{itemize}
  \item the function $\textit{drop} : \Nat \rightarrow [a] \rightarrow [a]$, which drops the first $n$ elements from a list, specializes to \textit{tail} by setting $n$ to $1$;
  \item the function $\textit{removeAt} : \Nat \rightarrow [a] \rightarrow [a]$, which removes an element from a list at index $i$, specializes to \textit{tail} by setting $i$ to $0$.
\end{itemize}
Neither \textit{drop} nor \textit{removeAt} can be implemented using the sketch $\lambda n.\ \textit{foldr}\ (f\ n)\ []$. There is, however, a way to implement \textit{drop} and \textit{removeAt} as a fold, using the sketch $\lambda n\ \textit{xs}.\ \textit{foldr}\ f\ e\ \textit{xs}\ n$, which instantiates \textit{foldr} such that the output type is $\Nat \rightarrow [a]$. See Section~\ref{sec:additional arguments} for a short discussion of these different interpretations.

\subsection{Sketching with Foldr}\label{sec:foldr}

\begin{figure}
  \centering
  \begin{tikzpicture}[auto,node distance=0mm],
    \node[spec,fill=type!25] (b1) {
      $\forall a.\ p : [F\,a] \rightarrow G\,a$
    };
    \node[left=of b1] {$\exists p.$};
    \node[right=of b1] (b12) {$\land$};
    \node[spec,right=of b12,fill=sketch!25] (b2) {
      $\exists f.\ p = \textit{foldr}\ f\ \outvar_0$
    };
    \node[right=of b2] (b23) {$\land$};
    \node[spec,right=of b23,fill=example!25] (b3) {
      $p\ [\invar_{n-1} \cdots \invar_0] \equiv \outvar_n$
    };
    \node[below=9mm of b2,xshift=2mm] (b45) {$\land$};
    \node[spec,left=of b45,fill=sketch+type!25] (b4) {
      $\forall a.\ f : F\,a \times G\,a \rightarrow G\,a$
    };
    \node[left=of b4] {$\exists f.\ \exists \outvar_1 \cdots \outvar_{n-1}.$};
    \node[spec,right=of b45,fill=example+sketch!25,stacked] (b5) {
      $f\ (\invar_i, \outvar_i) \equiv \outvar_{i+1}$
    };
    \node[right=1.5mm of b5,yshift=-1.5mm] {$_{0 \leq i < n}$};
    \node[spec,below=9mm of b45,xshift=2mm,fill=type+example+sketch!25,stacked] (b6) {
      $\llangle \shape{f}, \pos{f} \rrangle\
      ((\shape{\invar}_i, \shape{\outvar}_i), (\pos{\invar}_i \oplus
      \pos{\outvar}_i)) \equiv (\shape{\outvar}_{i+1}, \pos{\outvar}_{i+1})$
    };
    \node[left=of b6] {$\exists (\shape{f},\pos{f}).\
      \exists (\shape{\outvar}_1,\pos{\outvar}_1) \cdots
      (\shape{\outvar}_{n-1},\pos{\outvar}_{n-1}).$};
    \node[right=1.5mm of b6,yshift=-1.5mm] {$_{0 \leq i < n}$};
    \begin{scope}[arr]
      \draw (b1.south) |- ($(b1)!0.5!(b4)$) -| (b4.north);
      \draw (b2.south) +(-3mm,0) |- ($(b2)!0.5!(b4)$) -| (b4.north);
      \draw (b2.south) +(3mm,0)  |- ($(b2)!0.5!(b5)$) -| (b5.north);
      \draw (b3.south) |- ($(b3)!0.5!(b5)$) -| (b5.north);
      \draw (b4.south) |- ($(b4)!0.5!(b6)$) -| (b6.north);
      \draw (b5.south) +(0,-0.6mm) |- ($(b5)!0.5!(b6)$) -| (b6.north);
    \end{scope}
  \end{tikzpicture}
  \caption{The pipeline for computing the realizability of a polymorphic program defined as a fold with a single input-output example.}
  \label{fig:foldr pipeline}
\end{figure}
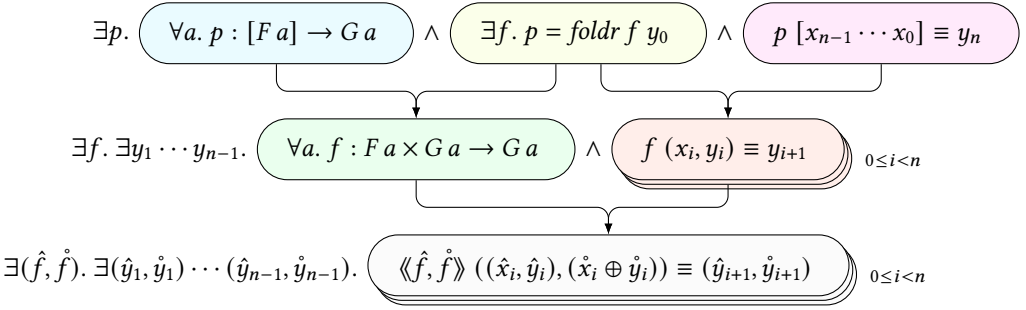

To reason about program traces generated by \textit{foldr}, we look at how \textit{foldr} iteratively builds up a result. When evaluating $\textit{foldr}\ (+)\ 1\ [2,3,4]$, two intermediate results are computed ($4 + 1 = 5$ and $3 + 5 = 8$) before the final result ($2 + 8 = 10$) is computed.\footnote{The exact order of evaluation depends on the evaluation strategy, but we can still reason about the intermediate results.} Note how each intermediate result is computed by combining the previous result with an element from the input list, going through the list from right to left (hence the name \textit{foldr}, which stands for \emph{right fold}). To reflect this, we write an input-output example for \textit{foldr} as follows, numbering the elements in the list from right to left:
\[
  \textit{foldr}\ f\ \outvar_0\ [\invar_{n-1},\dots,\invar_0] \equiv \outvar_n
\]
The argument $\outvar_0$ is the initial result and the output $\outvar_n$ is the final result. There are $n-1$ intermediate results ($\outvar_1, \dots, \outvar_{n-1})$, which we can make explicit:
\[
  \begin{array}{rl}
  \exists \outvar_1 \cdots \outvar_{n-1}.\ 
  &\outvar_1 \equiv f\ (\invar_0, \outvar_0)\\
  \land&\outvar_2 \equiv f\ (\invar_1, \outvar_1)\\
  &\vdots\\
  \land&\outvar_n \equiv f\ (\invar_{n-1},\outvar_{n-1})
  \end{array}
\]
This allows us to concisely define the input-output example on our fold as a set of input-output examples on $f$:
\[
  \exists \outvar_1 \cdots \outvar_{n-1}.\ 
  \bigwedge_{0 \leq i < n} f\ (\invar_i, \outvar_i) \equiv \outvar_{i+1}
\]
Along with the inferred type of $f$, these input-output examples are translated to the container setting. The resulting pipeline for programs defined as a fold is shown in Figure~\ref{fig:foldr pipeline}.

\paragraph*{Soundness}
Using the same argument as in Section~\ref{sec:map}, reasoning about the realizability of a program as a fold is sound, as all the steps shown in Figure~\ref{fig:foldr pipeline} are valid rewritings.

\paragraph*{Completeness}
When considering only a single input-output example, reasoning about the realizability of a program as a fold is \emph{not} complete. To see why this is the case, let us write out the final equation of Figure~\ref{fig:foldr pipeline} to separate the \textit{shape} and \textit{position} constraints:
\begin{equation} \label{eq:foldr separated}
  \bigwedge_{0 \leq i < n}\;\;
  \underbrace{
  \shape{f}\ (\shape{\invar}_i,\shape{\outvar}_i) \equiv
  \shape{\outvar}_{i+1}}_{\textit{shape}}
  \;\;\land\;\;
  \underbrace{
  \forall \outpos.\ (\pos{\invar}_i \oplus
  \pos{\outvar}_i)\ (\pos{f}\ \outpos)
  \equiv \pos{\outvar}_{i+1}\ \outpos}_{\textit{position}}
\end{equation}
Note that the values of the intermediate results $\outvar_1,\dots,\outvar_{n-1}$ (and therefore their shapes $\shape{\outvar}_1,\dots,\shape{\outvar}_{n-1}$) are not known. This means that we do not know the exact types of $\outpos$ and cannot simply eliminate the universal quantifications by enumeration as we did in Equation~\ref{eq:small morphism realizability}. In other words, the realizability of an input-output trace cannot be expressed in a quantifier-free logic.

If, however, all recursive calls are represented in the set of input-output examples (a property known as \emph{trace completeness}~\citep{osera_2015}), the values of the intermediate results would be known and all universal quantifiers could be eliminated.
Moreover, since the types of universally quantified variables are only determined by the \emph{shapes} of intermediate results, we can relax trace completeness to only require that the shape component of each recursive call is represented in the example set. Such an example set is called \emph{trace complete modulo parametricity} (or \emph{shape complete}). For example, the following set of input-output examples is shape complete:
\begin{center}
\begin{tikzpicture}[auto]
  \node[spec,fill=example!25] {
    $\textit{tail}\ [A,B,C] \equiv [B,C] \;\land\;
    \textit{tail}\ [1,2] \equiv [2] \;\land\;
    \textit{tail}\ [\textit{true}] \equiv []$
  };
\end{tikzpicture}
\end{center}
To conclude, reasoning about the realizability of a program as a fold is complete if $F$ and $G$ are small container functors and the set of input-output examples is shape complete.

\subsection{A Note on Scoping}\label{scoping-problems}

In the previous sections, we have shown how to rephrase the question of whether a function is a fold as a realizability problem. We have been very careful to state that a fold can be defined as $\textit{foldr}\ f\ e$, and not just defined ``in terms of \textit{foldr}". As it turns out, any function on lists can be defined in terms of \textit{foldr}! Take, for example, the following implementation for \textit{tail}:
\[\arraycolsep=1.4pt
\begin{array}{l}
  \textit{tail} :: \forall a.\ [a] \rightarrow [a] \\
  \textit{tail}\ \textit{xs} =
  \textit{foldr}\ (\lambda x\ r.\ \textit{tail'}\ \textit{xs})\ []\ \textit{xs} \\
  \begin{array}{rlcl}
    \quad \textbf{where}
      & \textit{tail'}\ [] &=& [] \\
      & \textit{tail'}\ (y : \textit{ys}) &=& \textit{ys}
  \end{array}
\end{array}
\]
More generally, any function \textit{h} on lists for which an implementation \textit{h'} exists can be implemented in terms of \textit{foldr} in at least two different ways:
\begin{align}
  \textit{h}\ \textit{xs} =&
    \ \textit{foldr}\ (\lambda x\ r.\ r)\ (\textit{h'}\ \textit{xs})\ \textit{xs}\\
  \textit{h}\ \textit{xs} =&
    \ \textit{foldr}\ (\lambda x\ r.\ \textit{h'}\ \textit{xs})\ (\textit{h'}\ [])\ \textit{xs}
\end{align}
The problem is that the algebra of a fold (represented in the function \textit{foldr} by the first two arguments) should be independent of the input list. Hence, to figure out if a function is a fold, we should make sure that \textit{xs} is not in scope of the arguments \textit{f} and \textit{e} when computing the realizability of $\textit{foldr}\ f\ e\ \textit{xs}$.

\section{Evaluation}

To test how well our technique is able to check for realizability, we will evaluate it on a benchmark of polymorphic Haskell functions, checking for each whether it can be implemented as a fold. To create this benchmark, we selected all functions from the Haskell prelude\footnote{\url{https://hackage.haskell.org/package/base-4.19.1.0/docs/Prelude.html}} that are natural transformations of type $\forall a.\ H\,a \times [F\,a] \rightarrow G\,a$, i.e.~any total polymorphic function taking at least a list as input. We made small changes to some functions to make them fit these criteria:
\begin{itemize}
  \item All functions with a \verb|Foldable| constraint are specialized to \verb|[]| (\verb|concat|, \verb|null|, \verb|length|).
  \item Partial functions that return a list return the empty list for invalid inputs (\verb|tail|, \verb|init|).
  \item Other partial functions have their return type wrapped in \verb|Maybe| (\verb|head|, \verb|last|, \verb|index|).
  \item Since there are two ways to make the function \verb|(++)| fit this scheme (i.e.~we can try to fold over either of its input lists), we made both ways explicit in \verb|append| and \verb|prepend|.
  \item Functions with multiple type parameters are specialized to the same type parameter (\verb|zip|, \verb|unzip|).
\end{itemize}
The functions are listed in Table~\ref{tab:benchmark}. For each function $p$ in the benchmark, we check if the sketch $\exists f.\ \textit{p}\ (x, \textit{ys}) = \textit{foldr}\ (f\ x)\ (e\ x)\ \textit{ys}$ is realizable:
\begin{center}
\begin{tikzpicture}[auto,node distance=0mm]
  \node[spec,fill=type!25] (b1) {
    $\forall a.\ \textit{p} : H\,a \times [F\,a] \rightarrow G\,a$
  };
  \node[right=of b1] (b12) {$\land$};
  \node[spec,right=of b12,fill=sketch!25] (b2) {
    $\exists f.\ \textit{p}\ (x, \textit{ys}) = \textit{foldr}\ (f\ x)\ (e\ x)\ \textit{ys}$
  };
  \node[right=of b2] (b23) {$\land$};
  \node[spec,right=of b23,fill=example!25] (b3) {
    $\matchlength{\kern-0.85mm\dots}{}$
  };
\end{tikzpicture}
\end{center}
Note how we use an extended version of the pipeline in Figure~\ref{fig:foldr pipeline} that takes an additional argument of type $H\,a$. This argument is passed to the function $f$ and allows us to check the realizability of functions taking additional arguments (beyond an input list). While there is no technical limitation that stops us from reasoning about multiple holes, for simplicity we assume that the base case $e\ x$ is given. As such, $e$ is not existentially quantified.

\begin{table}[t]
  \centering
  \caption{Benchmark computing the realizability of Haskell prelude functions as folds. For each function, its type is given and whether it is a fold. The running time with shape complete (SC) and shape incomplete (SI) example sets is shown in milliseconds.}
  \begin{tabular}{ l l | c | r | r }

    \verb|name| & \verb|:: type| & fold? & SC & SI \\

    \hline

    \verb|null|    & \verb|:: [a] -> Bool|              & \yes &  106 &   98 \\
    \verb|length|  & \verb|:: [a] -> Int|               & \yes &  110 &   99 \\
    \verb|head|    & \verb|:: [a] -> Maybe a|           & \yes &  123 &  107 \\
    \verb|last|    & \verb|:: [a] -> Maybe a|           & \yes &  121 &  107 \\
    \verb|tail|    & \verb|:: [a] -> [a]|               & \no  &  120 &  130 \\
    \verb|init|    & \verb|:: [a] -> [a]|               & \no  &  117 &  123 \\
    \verb|reverse| & \verb|:: [a] -> [a]|               & \yes &  157 &  144 \\
    \verb|index|   & \verb|:: Int -> [a] -> Maybe a|    & \no  &  131 &  132 \\
    \verb|drop|    & \verb|:: Int -> [a] -> [a]|        & \no  &  140 &  149 \\
    \verb|take|    & \verb|:: Int -> [a] -> [a]|        & \yes &  197 &  217 \\
    \verb|splitAt| & \verb|:: Int -> [a] -> ([a], [a])| & \yes &  566 &  545 \\
    \verb|append|  & \verb|:: [a] -> [a] -> [a]|        & \yes &  238 &  275 \\
    \verb|prepend| & \verb|:: [a] -> [a] -> [a]|        & \yes &  237 &  240 \\
    \verb|zip| & \verb|:: [a] -> [a] -> [(a, a)]| & \yes &  223\opt&  234\opt\\
    \verb|unzip| & \verb|:: [(a, a)] -> ([a], [a])| & \yes &  228\opt&$\bot$\\
    \verb|concat| & \verb|:: [[a]] -> [a]| & \yes &  247\opt&  318\opt\\

  \end{tabular}
  \label{tab:benchmark}
\end{table}

\subsection{Additional Arguments}\label{sec:additional arguments}

The functions \verb|index|, \verb|drop|, \verb|take|, and \verb|splitAt| take an integer as an argument in addition to the input list. There are multiple ways to interpret these functions as natural transformations between container functors. For example, for the function \verb|drop|, we can either
\begin{itemize}
  \item choose $H\,a = \textit{Int}$, $F\,a = a$, and $G\,a = [a]$, keeping the integer argument constant throughout the fold;
  \item or choose $H\,a = ()$, $F\,a = a$, and $G\,a = \textit{Int} \rightarrow [a]$, allowing a different integer to be passed to recursive calls.
\end{itemize}
In the first interpretation, \verb|drop| is \emph{not} a fold, but in the second interpretation it is:
\begin{verbatim}
  drop :: [a] -> Int -> [a]
  drop = foldr f (const [])
    where f x r i = if i > 0 then r (i - 1) else x : r i
\end{verbatim}
We only consider the first interpretation in our benchmark, since the functor $\textit{Int} \rightarrow [ - ]$ is \emph{not} a small container, as there are an infinite number of positions for every shape. The same approach is taken for all functions that take an extra argument. This includes \verb|index|, \verb|take|, and \verb|splitAt|, but also \verb|append|, \verb|prepend|, and \verb|zip|. In general, any additional arguments can either be kept constant or not. Unless the argument type has a finite number of values, the resulting container functors are \emph{not} small.

\subsection{Input-Output Examples}

Each function is tested on a shape complete example set curated by the author (each consisting of 4 to 10 input-output examples), as well as a shape incomplete example set constructed by removing every other example from the shape complete example set.

\subsection{Running Times}

For each function, we automatically prove its (un)realizability as a fold with shape complete (SC) and shape incomplete (SI) example sets. Each proof is performed 10 times and the average results are shown in Table~\ref{tab:benchmark}.

Interestingly, apart from \verb|unzip|, there is no significant difference in runtime between proofs with shape complete and shape incomplete example sets. Some proofs perform slightly faster with shape incomplete example sets. This could be explained by a naive choice we made in the implementation, which seems to prevent the solver from effectively making use of the guarantees that shape completeness provides (see Section~\ref{sec:efficient containers} for a more in-depth discussion). As a result, those shape incomplete example sets seem to perform better simply because they have fewer input-output examples, and thus fewer constraints.

With the exception of \verb|unzip| with shape incomplete examples, all automated proofs finish in less than a second. These running times are fast enough for automated program analysis and feedback, but not yet for extensive pruning during synthesis. However, our tool could be used for top-level pruning, where realizability is only checked at the start of a synthesis problem to figure out the right recursion scheme.

\subsection{Naive Container Translations}

The return type of the functions \verb|zip|, \verb|unzip|, and \verb|splitAt| uses a product. When using a naive translation to container functors using Definition~\ref{def:product container}, computing the realizability of these functions results in a timeout (marked \textdagger). Our tool uses a more efficient translation described in more depth in Section~\ref{sec:efficient containers}.

\section{Implementation}

We have implemented a tool for reasoning about the realizability of polymorphic programs. It includes the pipelines from Figures~\ref{fig:map pipeline} and~\ref{fig:foldr pipeline} and provides the basics for defining more realizability problems. The tool is implemented in Haskell, as it allows us to reason about actual Haskell functions and values, rather than describe an intermediate language to reason about. For translating to SMT-LIB, we use the SBV library,\footnote{\url{https://leventerkok.github.io/sbv/}} a well-documented library with many features that provides a statically typed API into SMT-LIB. We use Z3~\citep{moura_2008} as the underlying solver.

\subsection{Encoding Containers}

To encode a functor \verb|f|, we have to associate it to its shape and position types. To do so, we define two type families \verb|Shape| and \verb|Position|.
\begin{verbatim}
  type family Shape    (f :: Type -> Type) :: Type
  type family Position (f :: Type -> Type) :: Type
\end{verbatim}
For example, to associate the list functor \verb|[]| to a shape and position (see Definition~\ref{def:list container}) we define:
\begin{verbatim}
  newtype Fin = Fin Nat deriving (Eq, Ord, Num)
  type instance Shape    [] = Nat
  type instance Position [] = Fin
\end{verbatim}
Note that Haskell does not support dependent types, so we overapproximate \verb|Fin| as a newtype over natural numbers. Additionally, SMT-LIB has no native support for $\Nat$ and $\Fin$, so we employ another type family \verb|Sym| (for \emph{sym}bolic representation) to associate our Haskell datatypes to types supported by SMT-LIB, along with a type class \verb|Encode| for encoding values. Natively supported types are exactly those for which the constraint \verb|SymVal| holds, which is exported by the SBV library.
\begin{verbatim}
  type family Sym (a :: Type) :: Type
  class SymVal (Sym a) => Encode a where
    encode :: a -> SMT a
\end{verbatim}
Both \verb|Nat| and \verb|Fin| are represented in the SMT solver as integers:
\begin{verbatim}
  type instance Sym Nat = Integer
  type instance Sym Fin = Integer
  instance Encode Nat where encode = fromIntegral
  instance Encode Fin where encode = fromIntegral
\end{verbatim}
Of course, these are overapproximations, which we will resolve by describing how the symbolic representations should be constrained in the SMT solver. A value of type \verb|Nat| is encoded as an integer $n$ constrained by $n \geq 0$ and an associated value of type \verb|Fin| as an integer $m$ constrained by $m \geq 0 \land m < n$, for some natural number $n$. We refer to \verb|Nat| (and other shape types) as refinement types and to \verb|Fin| (and other position types) as dependent types.

The type class \verb|Ref| describes how the symbolic representation of a refinement type is constrained in terms of a function \verb|refine| returning a symbolic boolean \verb|SBool|. Note that when calling \verb|refine|, we have to disambiguate between multiple refinement types that have the same symbolic representation. To do so, we use visible dependent quantification (denoted \verb|forall a ->|) as introduced by the \verb|RequiredTypeArguments| extension,\footnote{\url{https://ghc.gitlab.haskell.org/ghc/doc/users_guide/exts/required_type_arguments.html}} to avoid having to use proxy arguments or ambiguous types.
\begin{verbatim}
  class Encode a => Ref a where
    refine :: forall a -> Sym a -> SBool
\end{verbatim}
By convention, operators that act on symbolic values start with a dot (\verb|.|).
\begin{verbatim}
  instance Ref Nat where
    refine _ n = n .>= 0
\end{verbatim}
For dependent types, we additionally define a type family \verb|Arg| to associate them with the argument type they depend on. The type class \verb|Dep| describes, for a dependent type, how its symbolic representation is constrained in terms of its argument.
\begin{verbatim}
  type family Arg (a :: Type) :: Type
  class (Encode a, Ref (Arg a)) => Dep a where
    depend :: forall a -> Sym (Arg a) -> Sym a -> SBool

  type instance Arg Fin = Nat
  instance Dep Fin where
    depend _ n m = m .>= 0 .&& m .< n
\end{verbatim}
Using \verb|Ref| and \verb|Dep|, we can define a type dependency between a refinement type \verb|t| and a dependent type \verb|u| by stating that the argument type of \verb|u| is equal to \verb|t|.
\begin{verbatim}
  type Dependent :: Type -> Type -> Constraint
  type Dependent t u = (Ref t, Dep u, Arg u ~ t)
\end{verbatim}
The type class \verb|Container| describes an isomorphism between a container functor \verb|f| and its extension, giving a dependency between its shape and position types.
\begin{verbatim}
  class Dependent (Shape f) (Position f) => Container f where
    toExtension   :: f a -> Extension f a
    fromExtension :: Extension f a -> f a
\end{verbatim}
As described in Section~\ref{sec:container functors}, the extension of a container is a shape along with a function mapping positions to elements. In our implementation, we choose to represent this function as a dictionary, to emphasize that we are dealing with \emph{small} containers.
\begin{verbatim}
  data Extension f a = Extension
    { shape    :: Shape f
    , position :: Map (Position f) a
    }
\end{verbatim}
Finally, we define \verb|Container []| by giving a translation to and from the container extension.
\begin{verbatim}
  instance Container [] where
    toExtension xs = Extension
      { shape    = genericLength xs
      , position = Map.fromList (zip [0..] xs)
      }
    fromExtension (Extension _ p) = Map.elems p
\end{verbatim}

\subsection{Encoding the Pipelines}

For each of the pipelines in Figures~\ref{fig:map pipeline} and~\ref{fig:foldr pipeline}, we introduce a function that takes a list of input-output examples and produces a set of constraints to be passed to the SMT solver. All inputs and outputs are translated to their container representation. For each existentially quantified variable in the final equation of the pipeline, an uninterpreted function is introduced. These uninterpreted functions are then constrained as in Equations~\ref{eq:morphism realizability} and~\ref{eq:foldr separated}, making sure to insert calls to \verb|refine| and \verb|depend| so that the functions are only constrained on their true domain. For example, a symbolic function of type $\textit{Int} \rightarrow \textit{Int}$ representing a shape morphism $\hat{f} : \Nat \rightarrow \Nat$ should not be constrained on negative inputs, and all its results should be nonnegative.

\subsection{Efficient Encodings of Position Sets} \label{sec:efficient containers}

In Section~\ref{sec:foldr} we describe how a shape complete example set leads to a quantifier-free formula because the quantified positions can be enumerated over. In our implementation, however, we have not made proper use of this observation. Rather, we simply describe how the shapes constrain the position types and leave it to the SMT solver to figure out that these constraints imply a finite set of positions. This approach turns out to be too naive, especially for position types that contain unions.\footnote{Both product and sum containers use unions in their position types.}

For example, the function \verb|splitAt| returns a pair of lists (i.e.~it is a natural transformation to the functor \verb|Product [] []|). Using the standard constructors for containers (as taken from~\citet{abbot_2005}), this corresponds to the container
\[
  ((n, m) : \Nat \times \Nat) \triangleright \Fin\ n + \Fin\ m
\]
The position type contains a union, and using this container to compute the realizability of \verb|splitAt| in terms of \verb|foldr| results in a timeout. However, an equivalent, more efficient container exists, namely the container
\[
  ((n, m) : \Nat \times \Nat) \triangleright \Fin\ (n + m)
\]
By using this optimized container representation, our tool can efficiently compute the realizability of \verb|splitAt| in terms of \verb|foldr|. Similarly, a list of pairs (the return type of \verb|zip| and \verb|unzip|) can efficiently be represented by the container $(n : \Nat) \triangleright \Fin\ 2n$.

In our tool, these efficient containers are defined ad-hoc, but they pave the way for a more general solution: every small container can be represented as $(s : S) \triangleright \Fin\ (f\ s)$, where $f$ is some function that returns the number of positions given a shape $s$ of type $S$. The problem is that $f$ may be difficult to turn into a constraint that is efficiently solved by an SMT solver. When the example set is shape complete, however, all shapes are known, making it possible to compute $f\ s$ before translating to SMT-LIB. This way, both shapes and positions could be represented in SMT-LIB as integers. We leave the exploration of this approach to future work.

\section{Related Work}

\subsection{Typed-Directed Program Synthesis}

Many program synthesizers use types to constrain the search space~\citep{katayama_2008, osera_2015, osera_2019, inala_2015, frankle_2016, polikarpova_2016, lubin_2020, feser_2015, koppel_2022, gissurarson_2023, mulleners_2023, lee_2023}. By inferring synthesis rules from typing rules, synthesized programs are type correct by construction~\citep{osera_2015, inala_2015}. Polymorphic types additionally constrain the search space, by restricting the production forms to those that make no assumptions about the types, allowing synthesizers to implicitly benefit from the abstractions provided by parametricity~\citep{reynolds_1983, wadler_1989}. This is emphasized when polymorphic types are combined with other specifications: often, parametricity can turn an otherwise partial specification into a total specification~\citep{frankle_2016, polikarpova_2016, osera_2019}. However, none of these synthesizers explicitly take the interaction between polymorphic types and other specifications into account.

\subsection{Reasoning About Polymorphic Functions}

Parametrically polymorphic functions behave the same regardless of how they are instantiated. This property is known as \emph{parametricity}. \citet{reynolds_1983} captures parametricity in his abstraction theorem, by giving a relational interpretation to types. \citet{wadler_1989} shows how this interpretation can be used to derive free theorems for functions based solely on their types. \citet{abbot_2003,abbot_2005} define the notion of \emph{containers} and show that any polymorphic function between strictly positive types corresponds to a morphism between containers. \citet{prince_2008} use morphisms between containers to prove properties about polymorphic functions. \citet{seidel_2010} extend this reasoning to ad-hoc polymorphic functions. \citet{bernardy_2010} show how properties of polymorphic functions can be faithfully tested on a single monomorphic instance. However, these techniques only reason about the correctness of complete programs, rather than the realizability of incomplete programs.

\subsection{Example Propagation}

Whereas input-output examples are commonly used to test the correctness of complete programs, they can often also be used to reason about the realizability of incomplete programs, a process known as \emph{example propagation}. This is of particular interest to top-down enumerative synthesizers, for pruning the search space. Example propagation is typically implemented using specific deduction rules~\citep{feser_2015,osera_2015,frankle_2016,osera_2019,hofmann_2010a}, live evaluation~\citep{omar_2019,lubin_2020,mulleners_2023}, or function inversion~\citep{lee_2023,teegen_2021}. In this paper, example propagation for \textit{map} and \textit{foldr} is defined in an ad-hoc manner, not unlike the deduction rules used by~\citet{feser_2015}.

\subsection{Unrealizability}

In recent years, a lot of advancements have been made in reasoning about the unrealizability of programs in the context of syntax-guided synthesis (\textsc{SyGuS})~\citep{hu_2018,hu_2019,hu_2020} and semantics-guided synthesis (\textsc{SemGuS})~\citep{kim_2021,kim_2023}. Typically, these techniques prove that a specification is unrealizable when restricted to a specific grammar. This allows synthesizers to incrementally extend the grammar until a minimal solution (in terms of program size) is found. A similar approach could be possible using unrealizability reasoning as described in this paper, where a synthesizer checks the realizability of a specification against a hierarchy of increasingly expressive recursion schemes.

A different approach to synthesis using unrealizability is taken by \citet{farzan_2022}. Their tool \textsc{Synduce} extends counterexample-guided inductive synthesis (CEGIS) with a dual inductive procedure for generating unrealizability witnesses.

\section{Conclusion}

We have shown how to automatically compute the realizability of polymorphic functions with input-output examples by interpreting those functions as container morphisms. This approach goes hand in hand with example propagation, a technique to compute the realizability of sketches with input-output examples. We have presented a general schema for computing the realizability of polymorphic functions with input-output examples and sketches, in the form a pipeline that combines type checking, example propagation, translation to containers, and SMT solving. Two concrete instantiations for this pipeline are presented for reasoning about programs defined using \textit{map} and \textit{foldr}. To support recursion schemes such as \textit{foldr}, we have extended our technique to reason not just about input-output examples, but input-output traces. Additionally, we have introduced the notion of shape completeness, a generalization of trace completeness~\citep{osera_2015} that takes parametricity into account, which is required to show that reasoning about folds is decidable.

\subsection{Future Work}

The examples in this paper focus primarily on functions acting on lists. It would, however, be interesting to explore more complex datatypes, such as binary trees. Our current technique is restricted to polymorphic functions between strictly positive, unary functors, as this allows an easy translation of inputs and outputs to the container setting as described by~\citet{abbot_2005}. We can relax these restrictions by extending the theory of containers. For example, switching from unary to $n$-ary containers would enable reasoning about multiple type variables. Supporting ad-hoc polymorphism and functors that are not strictly positive would require more complicated extensions to containers, such as described by~\citet{seidel_2010} and~\citet{bernardy_2010}. Alternatively, future work could explore the possibility of expressing the realizability of polymorphic programs directly in terms of relational parametricity~\citep{reynolds_1983}, rather than relying on the abstraction of container functors.

Our current approach to sketching is rather ad-hoc, describing only how to reason about examples propagated through \textit{map} and \textit{foldr}. While it is possible to add support for more such combinators, it would be interesting to take a more general approach to sketching, such as live bidirectional evaluation~\citep{lubin_2020} or automated function inversion~\citep{teegen_2021}.

Another avenue for future work is to explore the different applications of realizability reasoning, including automated program analysis in IDEs, automated feedback in programming tutors, and pruning in program synthesis.

\appendix

\section*{Data-Availability Statement}

All code is available on Zenodo~\citep{icfp_24_artifact} for reproduction and on GitHub\footnote{\url{https://github.com/NiekM/parametrickery.haskell}} for reuse.

\begin{acks}
We would like to thank Koen Claessen for helping work out the initial idea; Wouter Swierstra for suggesting the connection to container functors; the members of the ST4LT and ST groups for their support and discussions; and anonymous reviewers for many helpful suggestions.
\end{acks}

\bibliographystyle{ACM-Reference-Format}
\bibliography{citations}


\begin{thebibliography}{37}


\ifx \showCODEN    \undefined \def \showCODEN     #1{\unskip}     \fi
\ifx \showDOI      \undefined \def \showDOI       #1{#1}\fi
\ifx \showISBNx    \undefined \def \showISBNx     #1{\unskip}     \fi
\ifx \showISBNxiii \undefined \def \showISBNxiii  #1{\unskip}     \fi
\ifx \showISSN     \undefined \def \showISSN      #1{\unskip}     \fi
\ifx \showLCCN     \undefined \def \showLCCN      #1{\unskip}     \fi
\ifx \shownote     \undefined \def \shownote      #1{#1}          \fi
\ifx \showarticletitle \undefined \def \showarticletitle #1{#1}   \fi
\ifx \showURL      \undefined \def \showURL       {\relax}        \fi
\providecommand\bibfield[2]{#2}
\providecommand\bibinfo[2]{#2}
\providecommand\natexlab[1]{#1}
\providecommand\showeprint[2][]{arXiv:#2}

\bibitem[\protect\citeauthoryear{Abbott, Altenkirch, and Ghani}{Abbott
  et~al\mbox{.}}{2003}]%
        {abbot_2003}
\bibfield{author}{\bibinfo{person}{Michael Abbott}, \bibinfo{person}{Thorsten
  Altenkirch}, {and} \bibinfo{person}{Neil Ghani}.}
  \bibinfo{year}{2003}\natexlab{}.
\newblock \showarticletitle{Categories of Containers}. In
  \bibinfo{booktitle}{\emph{Proceedings of the 6th International Conference on
  Foundations of Software Science and Computation Structures and Joint European
  Conference on Theory and Practice of Software}} (Warsaw, Poland)
  \emph{(\bibinfo{series}{FOSSACS'03/ETAPS'03})}. \bibinfo{pages}{23–38}.
\newblock
\showISBNx{3540008977}
\urldef\tempurl%
\url{https://doi.org/10.1007/3-540-36576-1_2}
\showDOI{\tempurl}


\bibitem[\protect\citeauthoryear{Abbott, Altenkirch, and Ghani}{Abbott
  et~al\mbox{.}}{2005}]%
        {abbot_2005}
\bibfield{author}{\bibinfo{person}{Michael Abbott}, \bibinfo{person}{Thorsten
  Altenkirch}, {and} \bibinfo{person}{Neil Ghani}.}
  \bibinfo{year}{2005}\natexlab{}.
\newblock \showarticletitle{Containers: Constructing Strictly Positive Types}.
\newblock \bibinfo{journal}{\emph{Theor. Comput. Sci.}} \bibinfo{volume}{342},
  \bibinfo{number}{1} (\bibinfo{date}{sep} \bibinfo{year}{2005}),
  \bibinfo{pages}{3–27}.
\newblock
\showISSN{0304-3975}
\urldef\tempurl%
\url{https://doi.org/10.1016/j.tcs.2005.06.002}
\showDOI{\tempurl}


\bibitem[\protect\citeauthoryear{Bernardy, Jansson, and Claessen}{Bernardy
  et~al\mbox{.}}{2010}]%
        {bernardy_2010}
\bibfield{author}{\bibinfo{person}{Jean-Philippe Bernardy},
  \bibinfo{person}{Patrik Jansson}, {and} \bibinfo{person}{Koen Claessen}.}
  \bibinfo{year}{2010}\natexlab{}.
\newblock \showarticletitle{Testing Polymorphic Properties}. In
  \bibinfo{booktitle}{\emph{Programming Languages and Systems}},
  \bibfield{editor}{\bibinfo{person}{Andrew~D. Gordon}} (Ed.).
  \bibinfo{publisher}{Springer Berlin Heidelberg}, \bibinfo{address}{Berlin,
  Heidelberg}, \bibinfo{pages}{125--144}.
\newblock
\showISBNx{978-3-642-11957-6}
\urldef\tempurl%
\url{https://doi.org/10.1007/978-3-642-11957-6_8}
\showDOI{\tempurl}


\bibitem[\protect\citeauthoryear{de~Moura and Bj{\o}rner}{de~Moura and
  Bj{\o}rner}{2008}]%
        {moura_2008}
\bibfield{author}{\bibinfo{person}{Leonardo de Moura} {and}
  \bibinfo{person}{Nikolaj Bj{\o}rner}.} \bibinfo{year}{2008}\natexlab{}.
\newblock \showarticletitle{Z3: An Efficient SMT Solver}. In
  \bibinfo{booktitle}{\emph{Tools and Algorithms for the Construction and
  Analysis of Systems}}, \bibfield{editor}{\bibinfo{person}{C.~R. Ramakrishnan}
  {and} \bibinfo{person}{Jakob Rehof}} (Eds.). \bibinfo{publisher}{Springer
  Berlin Heidelberg}, \bibinfo{address}{Berlin, Heidelberg},
  \bibinfo{pages}{337--340}.
\newblock
\showISBNx{978-3-540-78800-3}
\urldef\tempurl%
\url{https://doi.org/10.1007/978-3-540-78800-3_24}
\showDOI{\tempurl}


\bibitem[\protect\citeauthoryear{Farzan, Lette, and Nicolet}{Farzan
  et~al\mbox{.}}{2022}]%
        {farzan_2022}
\bibfield{author}{\bibinfo{person}{Azadeh Farzan}, \bibinfo{person}{Danya
  Lette}, {and} \bibinfo{person}{Victor Nicolet}.}
  \bibinfo{year}{2022}\natexlab{}.
\newblock \showarticletitle{Recursion synthesis with unrealizability
  witnesses}. In \bibinfo{booktitle}{\emph{Proceedings of the 43rd ACM SIGPLAN
  International Conference on Programming Language Design and Implementation}}
  (San Diego, CA, USA) \emph{(\bibinfo{series}{PLDI 2022})}.
  \bibinfo{publisher}{Association for Computing Machinery},
  \bibinfo{address}{New York, NY, USA}, \bibinfo{pages}{244–259}.
\newblock
\showISBNx{9781450392655}
\urldef\tempurl%
\url{https://doi.org/10.1145/3519939.3523726}
\showDOI{\tempurl}


\bibitem[\protect\citeauthoryear{Felleisen, Findler, Flatt, and
  Krishnamurthi}{Felleisen et~al\mbox{.}}{2018}]%
        {felleisen_2018}
\bibfield{author}{\bibinfo{person}{Matthias Felleisen},
  \bibinfo{person}{Robert~Bruce Findler}, \bibinfo{person}{Matthew Flatt},
  {and} \bibinfo{person}{Shriram Krishnamurthi}.}
  \bibinfo{year}{2018}\natexlab{}.
\newblock \bibinfo{booktitle}{\emph{How to Design Programs: An Introduction to
  Programming and Computing}}.
\newblock \bibinfo{publisher}{The MIT Press}.
\newblock
\showISBNx{0262534800}


\bibitem[\protect\citeauthoryear{Feser, Chaudhuri, and Dillig}{Feser
  et~al\mbox{.}}{2015}]%
        {feser_2015}
\bibfield{author}{\bibinfo{person}{John~K. Feser}, \bibinfo{person}{Swarat
  Chaudhuri}, {and} \bibinfo{person}{Isil Dillig}.}
  \bibinfo{year}{2015}\natexlab{}.
\newblock \showarticletitle{Synthesizing data structure transformations from
  input-output examples}.
\newblock \bibinfo{journal}{\emph{ACM SIGPLAN Notices}} \bibinfo{volume}{50},
  \bibinfo{number}{6} (\bibinfo{date}{Aug.} \bibinfo{year}{2015}),
  \bibinfo{pages}{229--239}.
\newblock
\showISSN{0362-1340, 1558-1160}
\urldef\tempurl%
\url{https://doi.org/10.1145/2813885.2737977}
\showDOI{\tempurl}


\bibitem[\protect\citeauthoryear{Frankle, Osera, Walker, and Zdancewic}{Frankle
  et~al\mbox{.}}{2016}]%
        {frankle_2016}
\bibfield{author}{\bibinfo{person}{Jonathan Frankle},
  \bibinfo{person}{Peter-Michael Osera}, \bibinfo{person}{David Walker}, {and}
  \bibinfo{person}{Steve Zdancewic}.} \bibinfo{year}{2016}\natexlab{}.
\newblock \showarticletitle{Example-Directed Synthesis: A Type-Theoretic
  Interpretation}.
\newblock \bibinfo{journal}{\emph{SIGPLAN Not.}} \bibinfo{volume}{51},
  \bibinfo{number}{1} (\bibinfo{date}{jan} \bibinfo{year}{2016}),
  \bibinfo{pages}{802–815}.
\newblock
\showISSN{0362-1340}
\urldef\tempurl%
\url{https://doi.org/10.1145/2914770.2837629}
\showDOI{\tempurl}


\bibitem[\protect\citeauthoryear{{GHC Team}}{{GHC Team}}{2023}]%
        {haskell_holes}
\bibfield{author}{\bibinfo{person}{{GHC Team}}.}
  \bibinfo{year}{2023}\natexlab{}.
\newblock \bibinfo{booktitle}{\emph{GHC 9.8.1 User's Guide}}.
\newblock
\urldef\tempurl%
\url{https://downloads.haskell.org/~ghc/9.8.1/docs/users_guide/exts/typed_holes.html}
\showURL{%
\tempurl}


\bibitem[\protect\citeauthoryear{Gibbons, Hutton, and Altenkirch}{Gibbons
  et~al\mbox{.}}{2001}]%
        {gibbons_2001}
\bibfield{author}{\bibinfo{person}{Jeremy Gibbons}, \bibinfo{person}{Graham
  Hutton}, {and} \bibinfo{person}{Thorsten Altenkirch}.}
  \bibinfo{year}{2001}\natexlab{}.
\newblock \showarticletitle{When is a function a fold or an unfold?}
\newblock \bibinfo{journal}{\emph{Electronic Notes in Theoretical Computer
  Science}} \bibinfo{volume}{44}, \bibinfo{number}{1} (\bibinfo{year}{2001}),
  \bibinfo{pages}{146--160}.
\newblock
\showISSN{1571-0661}
\urldef\tempurl%
\url{https://doi.org/10.1016/S1571-0661(04)80906-X}
\showDOI{\tempurl}
\newblock
\shownote{CMCS 2001, Coalgebraic Methods in Computer Science (a Satellite Event
  of ETAPS 2001)}.


\bibitem[\protect\citeauthoryear{Gissurarson}{Gissurarson}{2018}]%
        {gissurarson_2018}
\bibfield{author}{\bibinfo{person}{Matth\'{\i}as~P\'{a}ll Gissurarson}.}
  \bibinfo{year}{2018}\natexlab{}.
\newblock \showarticletitle{Suggesting valid hole fits for typed-holes
  (experience report)}.
\newblock \bibinfo{journal}{\emph{SIGPLAN Not.}} \bibinfo{volume}{53},
  \bibinfo{number}{7} (\bibinfo{date}{sep} \bibinfo{year}{2018}),
  \bibinfo{pages}{179–185}.
\newblock
\showISSN{0362-1340}
\urldef\tempurl%
\url{https://doi.org/10.1145/3299711.3242760}
\showDOI{\tempurl}


\bibitem[\protect\citeauthoryear{Gissurarson, Roque, and Koppel}{Gissurarson
  et~al\mbox{.}}{2023}]%
        {gissurarson_2023}
\bibfield{author}{\bibinfo{person}{Matth{\'{\i}}as~P{\'{a}}ll Gissurarson},
  \bibinfo{person}{Diego Roque}, {and} \bibinfo{person}{James Koppel}.}
  \bibinfo{year}{2023}\natexlab{}.
\newblock \showarticletitle{Spectacular: Finding Laws from 25 Trillion Terms}.
  In \bibinfo{booktitle}{\emph{{IEEE} Conference on Software Testing,
  Verification and Validation, {ICST} 2023, Dublin, Ireland, April 16-20,
  2023}}. \bibinfo{publisher}{{IEEE}}, \bibinfo{pages}{293--304}.
\newblock
\urldef\tempurl%
\url{https://doi.org/10.1109/ICST57152.2023.00035}
\showDOI{\tempurl}


\bibitem[\protect\citeauthoryear{Hofmann}{Hofmann}{2010}]%
        {hofmann_2010b}
\bibfield{author}{\bibinfo{person}{Martin Hofmann}.}
  \bibinfo{year}{2010}\natexlab{}.
\newblock \showarticletitle{Data-Driven Detection of Catamorphisms - Towards
  Problem Specific Use of Program Schemes for Inductive Program Synthesis}. In
  \bibinfo{booktitle}{\emph{Preproceedings of the 22nd Symposium on
  Implementation and Application of Functional Languages (IFL 2010) ; Alphen
  aan den Rijn, 1. - 3. Sept. 2010}},
  \bibfield{editor}{\bibinfo{person}{Jurriaan Hage}} (Ed.).
  \bibinfo{address}{Utrecht}, \bibinfo{pages}{25 -- 39}.
\newblock
\newblock
\shownote{\url{https://fis.uni-bamberg.de/handle/uniba/3953}}.


\bibitem[\protect\citeauthoryear{Hofmann and Kitzelmann}{Hofmann and
  Kitzelmann}{2010}]%
        {hofmann_2010a}
\bibfield{author}{\bibinfo{person}{Martin Hofmann} {and}
  \bibinfo{person}{Emanuel Kitzelmann}.} \bibinfo{year}{2010}\natexlab{}.
\newblock \showarticletitle{I/O Guided Detection of List Catamorphisms: Towards
  Problem Specific Use of Program Templates in IP}. In
  \bibinfo{booktitle}{\emph{Proceedings of the 2010 ACM SIGPLAN Workshop on
  Partial Evaluation and Program Manipulation}} (Madrid, Spain)
  \emph{(\bibinfo{series}{PEPM '10})}. \bibinfo{publisher}{Association for
  Computing Machinery}, \bibinfo{address}{New York, NY, USA},
  \bibinfo{pages}{93–100}.
\newblock
\showISBNx{9781605587271}
\urldef\tempurl%
\url{https://doi.org/10.1145/1706356.1706375}
\showDOI{\tempurl}


\bibitem[\protect\citeauthoryear{Hu, Breck, Cyphert, D'Antoni, and Reps}{Hu
  et~al\mbox{.}}{2019}]%
        {hu_2019}
\bibfield{author}{\bibinfo{person}{Qinheping Hu}, \bibinfo{person}{Jason
  Breck}, \bibinfo{person}{John Cyphert}, \bibinfo{person}{Loris D'Antoni},
  {and} \bibinfo{person}{Thomas Reps}.} \bibinfo{year}{2019}\natexlab{}.
\newblock \showarticletitle{Proving Unrealizability for Syntax-Guided
  Synthesis}. In \bibinfo{booktitle}{\emph{Computer Aided Verification}},
  \bibfield{editor}{\bibinfo{person}{Isil Dillig} {and} \bibinfo{person}{Serdar
  Tasiran}} (Eds.). \bibinfo{publisher}{Springer International Publishing},
  \bibinfo{address}{Cham}, \bibinfo{pages}{335--352}.
\newblock
\showISBNx{978-3-030-25540-4}
\urldef\tempurl%
\url{https://doi.org/10.1007/978-3-030-25540-4_18}
\showDOI{\tempurl}


\bibitem[\protect\citeauthoryear{Hu, Cyphert, D'Antoni, and Reps}{Hu
  et~al\mbox{.}}{2020}]%
        {hu_2020}
\bibfield{author}{\bibinfo{person}{Qinheping Hu}, \bibinfo{person}{John
  Cyphert}, \bibinfo{person}{Loris D'Antoni}, {and} \bibinfo{person}{Thomas
  Reps}.} \bibinfo{year}{2020}\natexlab{}.
\newblock \showarticletitle{Exact and approximate methods for proving
  unrealizability of syntax-guided synthesis problems}. In
  \bibinfo{booktitle}{\emph{Proceedings of the 41st ACM SIGPLAN Conference on
  Programming Language Design and Implementation}} (London, UK)
  \emph{(\bibinfo{series}{PLDI 2020})}. \bibinfo{publisher}{Association for
  Computing Machinery}, \bibinfo{address}{New York, NY, USA},
  \bibinfo{pages}{1128–1142}.
\newblock
\showISBNx{9781450376136}
\urldef\tempurl%
\url{https://doi.org/10.1145/3385412.3385979}
\showDOI{\tempurl}


\bibitem[\protect\citeauthoryear{Hu and D'Antoni}{Hu and D'Antoni}{2018}]%
        {hu_2018}
\bibfield{author}{\bibinfo{person}{Qinheping Hu} {and} \bibinfo{person}{Loris
  D'Antoni}.} \bibinfo{year}{2018}\natexlab{}.
\newblock \showarticletitle{Syntax-Guided Synthesis with Quantitative Syntactic
  Objectives}. In \bibinfo{booktitle}{\emph{Computer Aided Verification}},
  \bibfield{editor}{\bibinfo{person}{Hana Chockler} {and}
  \bibinfo{person}{Georg Weissenbacher}} (Eds.). \bibinfo{publisher}{Springer
  International Publishing}, \bibinfo{address}{Cham},
  \bibinfo{pages}{386--403}.
\newblock
\showISBNx{978-3-319-96145-3}
\urldef\tempurl%
\url{https://doi.org/10.1007/978-3-319-96145-3_21}
\showDOI{\tempurl}


\bibitem[\protect\citeauthoryear{Inala, Qiu, Lerner, and Solar{-}Lezama}{Inala
  et~al\mbox{.}}{2015}]%
        {inala_2015}
\bibfield{author}{\bibinfo{person}{Jeevana~Priya Inala},
  \bibinfo{person}{Xiaokang Qiu}, \bibinfo{person}{Benjamin~S. Lerner}, {and}
  \bibinfo{person}{Armando Solar{-}Lezama}.} \bibinfo{year}{2015}\natexlab{}.
\newblock \showarticletitle{Type Assisted Synthesis of Recursive Transformers
  on Algebraic Data Types}.
\newblock \bibinfo{journal}{\emph{CoRR}}  \bibinfo{volume}{abs/1507.05527}
  (\bibinfo{year}{2015}).
\newblock
\showeprint[arXiv]{1507.05527}
\urldef\tempurl%
\url{http://arxiv.org/abs/1507.05527}
\showURL{%
\tempurl}


\bibitem[\protect\citeauthoryear{Katayama}{Katayama}{2008}]%
        {katayama_2008}
\bibfield{author}{\bibinfo{person}{Susumu Katayama}.}
  \bibinfo{year}{2008}\natexlab{}.
\newblock \showarticletitle{Efficient Exhaustive Generation of Functional
  Programs Using Monte-Carlo Search with Iterative Deepening}. In
  \bibinfo{booktitle}{\emph{PRICAI 2008: Trends in Artificial Intelligence}},
  \bibfield{editor}{\bibinfo{person}{Tu-Bao Ho} {and} \bibinfo{person}{Zhi-Hua
  Zhou}} (Eds.). \bibinfo{publisher}{Springer Berlin Heidelberg},
  \bibinfo{pages}{199--210}.
\newblock
\showISBNx{978-3-540-89197-0}
\urldef\tempurl%
\url{https://doi.org/10.1007/978-3-540-89197-0_21}
\showDOI{\tempurl}


\bibitem[\protect\citeauthoryear{Kim, D'Antoni, and Reps}{Kim
  et~al\mbox{.}}{2023}]%
        {kim_2023}
\bibfield{author}{\bibinfo{person}{Jinwoo Kim}, \bibinfo{person}{Loris
  D'Antoni}, {and} \bibinfo{person}{Thomas Reps}.}
  \bibinfo{year}{2023}\natexlab{}.
\newblock \showarticletitle{Unrealizability Logic}.
\newblock \bibinfo{journal}{\emph{Proc. ACM Program. Lang.}}
  \bibinfo{volume}{7}, \bibinfo{number}{POPL}, Article \bibinfo{articleno}{23}
  (\bibinfo{date}{jan} \bibinfo{year}{2023}), \bibinfo{numpages}{30}~pages.
\newblock
\urldef\tempurl%
\url{https://doi.org/10.1145/3571216}
\showDOI{\tempurl}


\bibitem[\protect\citeauthoryear{Kim, Hu, D'Antoni, and Reps}{Kim
  et~al\mbox{.}}{2021}]%
        {kim_2021}
\bibfield{author}{\bibinfo{person}{Jinwoo Kim}, \bibinfo{person}{Qinheping Hu},
  \bibinfo{person}{Loris D'Antoni}, {and} \bibinfo{person}{Thomas Reps}.}
  \bibinfo{year}{2021}\natexlab{}.
\newblock \showarticletitle{Semantics-guided synthesis}.
\newblock \bibinfo{journal}{\emph{Proc. ACM Program. Lang.}}
  \bibinfo{volume}{5}, \bibinfo{number}{POPL}, Article \bibinfo{articleno}{30}
  (\bibinfo{date}{jan} \bibinfo{year}{2021}), \bibinfo{numpages}{32}~pages.
\newblock
\urldef\tempurl%
\url{https://doi.org/10.1145/3434311}
\showDOI{\tempurl}


\bibitem[\protect\citeauthoryear{Koppel, Guo, de~Vries, Solar-Lezama, and
  Polikarpova}{Koppel et~al\mbox{.}}{2022}]%
        {koppel_2022}
\bibfield{author}{\bibinfo{person}{James Koppel}, \bibinfo{person}{Zheng Guo},
  \bibinfo{person}{Edsko de Vries}, \bibinfo{person}{Armando Solar-Lezama},
  {and} \bibinfo{person}{Nadia Polikarpova}.} \bibinfo{year}{2022}\natexlab{}.
\newblock \showarticletitle{Searching Entangled Program Spaces}.
\newblock \bibinfo{journal}{\emph{Proc. ACM Program. Lang.}}
  \bibinfo{volume}{6}, \bibinfo{number}{{ICFP}}, Article
  \bibinfo{articleno}{91} (\bibinfo{date}{Aug} \bibinfo{year}{2022}),
  \bibinfo{numpages}{29}~pages.
\newblock
\urldef\tempurl%
\url{https://doi.org/10.1145/3547622}
\showDOI{\tempurl}


\bibitem[\protect\citeauthoryear{Lee and Cho}{Lee and Cho}{2023}]%
        {lee_2023}
\bibfield{author}{\bibinfo{person}{Woosuk Lee} {and} \bibinfo{person}{Hangyeol
  Cho}.} \bibinfo{year}{2023}\natexlab{}.
\newblock \showarticletitle{Inductive Synthesis of Structurally Recursive
  Functional Programs from Non-Recursive Expressions}.
\newblock \bibinfo{journal}{\emph{Proc. ACM Program. Lang.}}
  \bibinfo{volume}{7}, \bibinfo{number}{POPL}, Article \bibinfo{articleno}{70}
  (\bibinfo{date}{jan} \bibinfo{year}{2023}), \bibinfo{numpages}{31}~pages.
\newblock
\urldef\tempurl%
\url{https://doi.org/10.1145/3571263}
\showDOI{\tempurl}


\bibitem[\protect\citeauthoryear{Lubin, Collins, Omar, and Chugh}{Lubin
  et~al\mbox{.}}{2020}]%
        {lubin_2020}
\bibfield{author}{\bibinfo{person}{Justin Lubin}, \bibinfo{person}{Nick
  Collins}, \bibinfo{person}{Cyrus Omar}, {and} \bibinfo{person}{Ravi Chugh}.}
  \bibinfo{year}{2020}\natexlab{}.
\newblock \showarticletitle{Program Sketching with Live Bidirectional
  Evaluation}.
\newblock \bibinfo{journal}{\emph{Proc. ACM Program. Lang.}}
  \bibinfo{volume}{4}, \bibinfo{number}{{ICFP}}, Article
  \bibinfo{articleno}{109} (\bibinfo{date}{Aug.} \bibinfo{year}{2020}),
  \bibinfo{numpages}{29}~pages.
\newblock
\urldef\tempurl%
\url{https://doi.org/10.1145/3408991}
\showDOI{\tempurl}


\bibitem[\protect\citeauthoryear{Mulleners}{Mulleners}{2024}]%
        {icfp_24_artifact}
\bibfield{author}{\bibinfo{person}{Niek Mulleners}.}
  \bibinfo{year}{2024}\natexlab{}.
\newblock \bibinfo{title}{Reproduction Package for the ICFP 2024 Article
  `Example-Based Reasoning about the Realizability of Polymorphic Programs'}.
\newblock \bibinfo{howpublished}{Zenodo}.
\newblock
\urldef\tempurl%
\url{https://doi.org/10.5281/zenodo.11470781}
\showDOI{\tempurl}


\bibitem[\protect\citeauthoryear{Mulleners, Jeuring, and Heeren}{Mulleners
  et~al\mbox{.}}{2023}]%
        {mulleners_2023}
\bibfield{author}{\bibinfo{person}{Niek Mulleners}, \bibinfo{person}{Johan
  Jeuring}, {and} \bibinfo{person}{Bastiaan Heeren}.}
  \bibinfo{year}{2023}\natexlab{}.
\newblock \showarticletitle{Program Synthesis Using Example Propagation}. In
  \bibinfo{booktitle}{\emph{Practical Aspects of Declarative Languages: 25th
  International Symposium, PADL 2023, Boston, MA, USA, January 16–17, 2023,
  Proceedings}} (Boston , MA, USA). \bibinfo{publisher}{Springer-Verlag},
  \bibinfo{address}{Berlin, Heidelberg}, \bibinfo{pages}{20–36}.
\newblock
\showISBNx{978-3-031-24840-5}
\urldef\tempurl%
\url{https://doi.org/10.1007/978-3-031-24841-2_2}
\showDOI{\tempurl}


\bibitem[\protect\citeauthoryear{Norell}{Norell}{2009}]%
        {norell_2009}
\bibfield{author}{\bibinfo{person}{Ulf Norell}.}
  \bibinfo{year}{2009}\natexlab{}.
\newblock \bibinfo{booktitle}{\emph{Dependently Typed Programming in Agda}}.
\newblock \bibinfo{publisher}{Springer Berlin Heidelberg},
  \bibinfo{address}{Berlin, Heidelberg}, \bibinfo{pages}{230--266}.
\newblock
\showISBNx{978-3-642-04652-0}
\urldef\tempurl%
\url{https://doi.org/10.1007/978-3-642-04652-0_5}
\showDOI{\tempurl}


\bibitem[\protect\citeauthoryear{Omar, Voysey, Chugh, and Hammer}{Omar
  et~al\mbox{.}}{2019}]%
        {omar_2019}
\bibfield{author}{\bibinfo{person}{Cyrus Omar}, \bibinfo{person}{Ian Voysey},
  \bibinfo{person}{Ravi Chugh}, {and} \bibinfo{person}{Matthew~A. Hammer}.}
  \bibinfo{year}{2019}\natexlab{}.
\newblock \showarticletitle{Live Functional Programming with Typed Holes}.
\newblock \bibinfo{journal}{\emph{Proc. ACM Program. Lang.}}
  \bibinfo{volume}{3}, \bibinfo{number}{{POPL}}, Article
  \bibinfo{articleno}{14} (\bibinfo{date}{Jan.} \bibinfo{year}{2019}),
  \bibinfo{numpages}{32}~pages.
\newblock
\urldef\tempurl%
\url{https://doi.org/10.1145/3290327}
\showDOI{\tempurl}


\bibitem[\protect\citeauthoryear{Osera}{Osera}{2019}]%
        {osera_2019}
\bibfield{author}{\bibinfo{person}{Peter-Michael Osera}.}
  \bibinfo{year}{2019}\natexlab{}.
\newblock \showarticletitle{Constraint-Based Type-Directed Program Synthesis}.
  In \bibinfo{booktitle}{\emph{Proceedings of the 4th ACM SIGPLAN International
  Workshop on Type-Driven Development}} (Berlin, Germany)
  \emph{(\bibinfo{series}{TyDe 2019})}. \bibinfo{publisher}{Association for
  Computing Machinery}, \bibinfo{address}{New York, NY, USA},
  \bibinfo{pages}{64–76}.
\newblock
\showISBNx{9781450368155}
\urldef\tempurl%
\url{https://doi.org/10.1145/3331554.3342608}
\showDOI{\tempurl}


\bibitem[\protect\citeauthoryear{Osera and Zdancewic}{Osera and
  Zdancewic}{2015}]%
        {osera_2015}
\bibfield{author}{\bibinfo{person}{Peter-Michael Osera} {and}
  \bibinfo{person}{Steve Zdancewic}.} \bibinfo{year}{2015}\natexlab{}.
\newblock \showarticletitle{Type-and-Example-Directed Program Synthesis}. In
  \bibinfo{booktitle}{\emph{Proceedings of the 36th ACM SIGPLAN Conference on
  Programming Language Design and Implementation}} (Portland, OR, USA)
  \emph{(\bibinfo{series}{PLDI '15})}. \bibinfo{publisher}{Association for
  Computing Machinery}, \bibinfo{address}{New York, NY, USA},
  \bibinfo{pages}{619–630}.
\newblock
\showISBNx{9781450334686}
\urldef\tempurl%
\url{https://doi.org/10.1145/2737924.2738007}
\showDOI{\tempurl}


\bibitem[\protect\citeauthoryear{Polikarpova, Kuraj, and
  Solar-Lezama}{Polikarpova et~al\mbox{.}}{2016}]%
        {polikarpova_2016}
\bibfield{author}{\bibinfo{person}{Nadia Polikarpova}, \bibinfo{person}{Ivan
  Kuraj}, {and} \bibinfo{person}{Armando Solar-Lezama}.}
  \bibinfo{year}{2016}\natexlab{}.
\newblock \showarticletitle{Program Synthesis from Polymorphic Refinement
  Types}. In \bibinfo{booktitle}{\emph{Proceedings of the 37th ACM SIGPLAN
  Conference on Programming Language Design and Implementation}} (Santa
  Barbara, CA, USA) \emph{(\bibinfo{series}{PLDI '16})}.
  \bibinfo{publisher}{Association for Computing Machinery},
  \bibinfo{address}{New York, NY, USA}, \bibinfo{pages}{522–538}.
\newblock
\showISBNx{9781450342612}
\urldef\tempurl%
\url{https://doi.org/10.1145/2908080.2908093}
\showDOI{\tempurl}


\bibitem[\protect\citeauthoryear{Prince, Ghani, and McBride}{Prince
  et~al\mbox{.}}{2008}]%
        {prince_2008}
\bibfield{author}{\bibinfo{person}{Rawle Prince}, \bibinfo{person}{Neil Ghani},
  {and} \bibinfo{person}{Conor McBride}.} \bibinfo{year}{2008}\natexlab{}.
\newblock \showarticletitle{Proving Properties about Lists Using Containers}.
  In \bibinfo{booktitle}{\emph{Proceedings of the 9th International Conference
  on Functional and Logic Programming}} (Ise, Japan)
  \emph{(\bibinfo{series}{FLOPS'08})}. \bibinfo{publisher}{Springer-Verlag},
  \bibinfo{address}{Berlin, Heidelberg}, \bibinfo{pages}{97–112}.
\newblock
\showISBNx{3540789685}
\urldef\tempurl%
\url{https://doi.org/10.1007/978-3-540-78969-7_9}
\showDOI{\tempurl}


\bibitem[\protect\citeauthoryear{Reynolds}{Reynolds}{1983}]%
        {reynolds_1983}
\bibfield{author}{\bibinfo{person}{John~C. Reynolds}.}
  \bibinfo{year}{1983}\natexlab{}.
\newblock \showarticletitle{Types, Abstraction and Parametric Polymorphism}. In
  \bibinfo{booktitle}{\emph{IFIP Congress}}.
\newblock


\bibitem[\protect\citeauthoryear{Seidel and Voigtl\"{a}nder}{Seidel and
  Voigtl\"{a}nder}{2010}]%
        {seidel_2010}
\bibfield{author}{\bibinfo{person}{Daniel Seidel} {and} \bibinfo{person}{Janis
  Voigtl\"{a}nder}.} \bibinfo{year}{2010}\natexlab{}.
\newblock \showarticletitle{Proving Properties about Functions on Lists
  Involving Element Tests}. In \bibinfo{booktitle}{\emph{Proceedings of the
  20th International Conference on Recent Trends in Algebraic Development
  Techniques}} (Etelsen, Germany) \emph{(\bibinfo{series}{WADT'10})}.
  \bibinfo{publisher}{Springer-Verlag}, \bibinfo{address}{Berlin, Heidelberg},
  \bibinfo{pages}{270–286}.
\newblock
\showISBNx{9783642284113}
\urldef\tempurl%
\url{https://doi.org/10.1007/978-3-642-28412-0_17}
\showDOI{\tempurl}


\bibitem[\protect\citeauthoryear{Teegen, Prott, and Bunkenburg}{Teegen
  et~al\mbox{.}}{2021}]%
        {teegen_2021}
\bibfield{author}{\bibinfo{person}{Finn Teegen}, \bibinfo{person}{Kai-Oliver
  Prott}, {and} \bibinfo{person}{Niels Bunkenburg}.}
  \bibinfo{year}{2021}\natexlab{}.
\newblock \showarticletitle{Haskell{$^{-1}$}: Automatic Function Inversion in
  Haskell}. In \bibinfo{booktitle}{\emph{Proceedings of the 14th ACM SIGPLAN
  International Symposium on Haskell}} (Virtual, Republic of Korea)
  \emph{(\bibinfo{series}{Haskell 2021})}. \bibinfo{publisher}{Association for
  Computing Machinery}, \bibinfo{address}{New York, NY, USA},
  \bibinfo{pages}{41–55}.
\newblock
\showISBNx{9781450386159}
\urldef\tempurl%
\url{https://doi.org/10.1145/3471874.3472982}
\showDOI{\tempurl}


\bibitem[\protect\citeauthoryear{Urzyczyn}{Urzyczyn}{1997}]%
        {urzyczyn_1997}
\bibfield{author}{\bibinfo{person}{Pawel Urzyczyn}.}
  \bibinfo{year}{1997}\natexlab{}.
\newblock \showarticletitle{Inhabitation in typed lambda-calculi (a syntactic
  approach)}. In \bibinfo{booktitle}{\emph{Typed Lambda Calculi and
  Applications}}, \bibfield{editor}{\bibinfo{person}{Philippe de~Groote} {and}
  \bibinfo{person}{J.~Roger~Hindley}} (Eds.). \bibinfo{publisher}{Springer
  Berlin Heidelberg}, \bibinfo{address}{Berlin, Heidelberg},
  \bibinfo{pages}{373--389}.
\newblock
\showISBNx{978-3-540-68438-1}
\urldef\tempurl%
\url{https://doi.org/10.1007/3-540-62688-3_47}
\showDOI{\tempurl}


\bibitem[\protect\citeauthoryear{Wadler}{Wadler}{1989}]%
        {wadler_1989}
\bibfield{author}{\bibinfo{person}{Philip Wadler}.}
  \bibinfo{year}{1989}\natexlab{}.
\newblock \showarticletitle{Theorems for Free!}. In
  \bibinfo{booktitle}{\emph{Proceedings of the Fourth International Conference
  on Functional Programming Languages and Computer Architecture}} (Imperial
  College, London, United Kingdom) \emph{(\bibinfo{series}{FPCA '89})}.
  \bibinfo{publisher}{Association for Computing Machinery},
  \bibinfo{address}{New York, NY, USA}, \bibinfo{pages}{347–359}.
\newblock
\showISBNx{0897913280}
\urldef\tempurl%
\url{https://doi.org/10.1145/99370.99404}
\showDOI{\tempurl}


\end{thebibliography}

\end{document}